\documentclass[12pt,preprint]{aastex}

\def\arcsecpoint{$''\!.$}
\def\deg{$^{\rm o}$}

\shortauthors{Kraemer et al.}
\shorttitle{Outflowing UV Absorbers in NGC 4051}

\begin{document}

\title{Observations of Outflowing UV Absorbers in NGC~4051 with
the Cosmic Origins Spectrograph\altaffilmark{1}}

\author{S.B. Kraemer\altaffilmark{2},
D.M. Crenshaw\altaffilmark{3},
J.P. Dunn\altaffilmark{4},
T.J. Turner\altaffilmark{5},
A.P. Lobban\altaffilmark{6},
L. Miller\altaffilmark{7},
J.N. Reeves\altaffilmark{6},
T.C. Fischer\altaffilmark{3}
and V. Braito\altaffilmark{8}}

\altaffiltext{1}{Based on observations made with the NASA/ESA Hubble Space 
Telescope, obtained at the Space Telescope Science Institute, which is 
operated by the Association of Universities for Research in Astronomy,
Inc. under NASA contract NAS 5-26555. These observations are associated with 
proposal 11834.}

\altaffiltext{2}{Institute for Astrophysics and Computational Sciences,
Department of Physics, The Catholic University of America, Washington, DC
20064, USA; steven.b.kraemer@nasa.gov}

\altaffiltext{3}{Department of Physics and Astronomy, Georgia State 
University, Astronomy Offices, One Park Place South SE, Suite 700,
Atlanta, GA 30303, USA}

\altaffiltext{4}{Department of Chemistry and Physics, Augusta State University,
2500 Walton Way, Augusta, Ga. 30904, USA}

\altaffiltext{5}{Department of Physics, University of Maryland Baltimore County,
Baltimore, MD 21250, USA}

\altaffiltext{6}{Astrophysics Group, School of Physical and Geographical
Sciences, Keele University, Keele, Staffordshire ST5 5BG, UK}

\altaffiltext{7}{Department of Physics, University of Oxford, Denys Wilkinson
Building, Keble Road, Oxford OX1 3RH, UK}

\altaffiltext{8}{Department of Physics and Astronomy, University of Leicester,
University Rd., Leicester, LE1 7RH, U.K.}
 
\begin{abstract}

We present new {\it Hubble Space Telescope (HST)}/Cosmic Origins Spectrograph
observations of the Narrow-Line Seyfert 1 galaxy NGC 4051. These data were
obtained as part of a coordinated observing program including X-ray observations
with the {\it Chandra}/High Energy Transmission Grating (HETG) Spectrometer and
{\it Suzaku}. We detected nine kinematic components of UV absorption, which were
 previously identified using the {\it HST}/Space Telescope Imaging Spectrograph.
None of the absorption components showed evidence for changes in column density
or profile within the $\sim$ 10 yr between the STIS and COS observations, which
we interpret as evidence of 1) saturation, for the stronger components, or 2)
very low densities, i.e., $n_{\rm H}$ $<$ 1 cm$^{-3}$, for the weaker
components. After applying a $+$200 km s$^{-1}$ offset to the HETG spectrum, we
found that the radial velocities of the UV absorbers lay within the O~VII
profile. Based on photoionization models, we suggest that, while UV components
2, 5 and 7 produce significant O~VII absorption, the bulk of the X-ray
absorption detected in the HETG analysis occurs in more highly ionized gas.
Moreover, the mass loss rate is dominated by high ionization gas which lacks a
significant UV footprint.

\end{abstract}


\section{Introduction}

According to the standard paradigm, Active Galactic Nuclei (AGN) are powered by
accretion of matter onto a supermassive black hole. Outflowing winds may arise
from an accretion disk surrounding the black hole (e.g., Rees 1987) or at larger
distances. As evidence for such winds, more than 50\% of Seyfert 1 galaxies,
relatively local ($z<0.1$), modest luminosity ($L_{bol} < 10^{45} {\rm erg}
s^{-1}$) AGN, show intrinsic X-ray and UV absorption (Crenshaw, Kraemer, \&
George 2003, and references therein), suggesting that the absorbers have global
covering factors $> 0.5$. Blue-shifted absorption lines in their UV (Crenshaw et
al. 1999), and X-ray (Kaastra et al. 2000; Kaspi et al. 2000) spectra reveal
significant outflow velocities (up to $-$4000 km s$^{-1}$; Dunn et al. 2007).
The inferred mass-loss rates are comparable to the accretion rates needed to
produce the observed luminosities of AGN. Hence, mass outflows are a critical
component in the structure, energetics, and evolution of AGN. Various
acceleration mechanisms have been proposed for these outflows, in particular
radiative driving (e.g., Murray et al. 1995), thermal winds (Begelman, McKee, \&
Shields 1983), and magneto-hydrodynamic flows (Blandford \& Payne 1982). 

Overall, there appears to be a one-to-one correspondence between X-ray and UV
absorption in Seyfert galaxies (Crenshaw et al. 2003; but, see Dunn et al.
2008). Nevertheless, the nature of the physical connection between the sources
of UV and X-ray absorption may vary among these objects and, indeed, among
different kinematic components in individual objects. While it has been
suggested that at least some of the UV absorption arises in the same gas as the
X-ray absorbers (e.g., Mathur, Elvis, \& Wilkes 1999), it is apparent that there
is a range of physical conditions within the absorbing gas (e.g., Kraemer et al.
2002; 2003; 2005; Gabel et al. 2005). In fact, there is strong evidence that the
X-ray absorbers in individual objects span a range in ionization (e.g.,
Steenbrugge et al. 2005; Turner et al. 2011). While there are cases in which the
UV absorption lines are simply the ``footprint'' of the X-ray absorbers (e.g.,
Kraemer et al. 2005), an interesting possibility is that they are condensations
in a more highly ionized medium (Kriss et al. 1996, 2000; Krolik \& Kriss 2001).
For example, UV and X-ray absorbers may be detected at the same radial
velocities, but there is evidence, such as lower line-of-sight covering factors
for the UV absorbers (Kraemer et al. 2006) or constraints based on
photo-ionization modeling, that they are distinct physical components. If these
components are indeed co-located, it raises questions about the stability of
lower-ionization condensations, which may not be in pressure equilibrium with
the surrounding medium (e.g. Gabel et al. 2005), and the overall dynamics of the
outflows, because the integrated cross-section for absorption of radiation is
significantly greater for lower ionization gas (eg., Arav, Li \& Begelman 1994).

NGC 4051 is a nearby (distance $=$ 15.2 Mpc, Russel 2004), narrow-line Seyfert 1
(NLSy1) galaxy (see Osterbrock \& Pogge 1985), as evidenced by the narrowness of
its H$\beta$ emission line profile, with a full width at half maximum (FWHM)
$\approx$ 1070 km s$^{-1}$ (Peterson et al. 2004). Via optical reverberation
mapping, Denney et al. (2009) determined the mass of the central black hole to
be 1.73$^{+0.55}_{-0.52} \times 10^{6}$ M$_\odot$. NLSy1s are thought to possess
relatively smaller black holes than broad-line Seyfert 1s of similar luminosity,
hence have higher relative mass-accretion rates and are radiating at or near the
Eddington limit (e.g., Mathur 2000). Interestingly, while there is evidence that
NGC 4051 is substantially sub-Eddington (see Wang \& Netzer 2003), hence may not
be a typical NLSy1 in that sense, it does exhibit the extreme X-ray variability
characteristic of the class (e.g., Turner et al. 1999; Leighly 1999). 

While observations such as those with {\it ROSAT} (Komossa \& Fink 1997) and
{\it ASCA} (Guainazzi et al. 1996) originally revealed the presence of intrinsic
X-ray absorption in NGC 4051, it has been the subject of several more recent
studies at higher spectral resolution, both in the X-ray and UV. Collinge et al.
(2001) observed NGC 4051 on 2000 April 24 -- 25 with the {\it Chandra} High
Energy Transmission Grating (HETG) Spectrometer, with a total exposure time of
$\sim$ 81.5 ksec, and on 2000 March 24 -- 25 with the {\it Hubble Space
Telescope (HST)}/Space Telescope Imaging Spectrograph (STIS). They found that
the UV absorption showed 9 distinct kinematic components, all seen in C~IV, N~V,
and several in Si~IV, Si~III, Si~II, and C~II, while the X-ray absorption
consisted of two systems. They suggested that lower velocity X-ray system, with
radial velocity $v_{r}$ $\sim -600$ km s$^{-1}$, may be connected to the UV
absorbers. In {\it Far Ultraviolet Spectroscopic Explorer (FUSE)} spectra,
obtained on 2002 March 29, 2003 January 18, and 2003 March 19, Kaspi et al.
(2004) detected O~VI, H~I Lyman, and C~III absorption consistent with the
components detected by Collinge et al. In their analysis of a 100 ksec XMM-{\it
Newton} observation from 2001 May 16-17, Krongold et al. (2007) found evidence
for two X-ray absorption components. Based on time variability and
photoionization modeling, they argued that both systems were within 3.5 lt-days
of the central source. Steenbrugge et al. (2009; hereafter S2009) analyzed
$\sim$ 200 ksec of {\it Chandra}/Low Energy Transmission Grating (LETG)
spectrometer data, obtained 2001 December 31 -- 2002 January 1 and 2003 July
23 -- 24. They modeled the X-ray absorption with four components, three spanning
the range of $-200$ km s$^{-1} \leq$ $v_{r}$ $\leq$ $-610$ km s$^{-1}$ and one
at $\sim$ $-4760$ km s$^{-1}$, which could possibly be identified with the high
velocity/high ionization absorber suggested by Collinge et al. (although that
system was at $v_{r}$ $\sim -2340$ km s$^{-1}$).  As with Krongold et al.,
Steenbrugge et al. constrained the radial distances of the X-ray components
based on time variability (we will discuss this in more detail in Section 5.).

We observed NGC 4051 for 300 ksec with {\it Chandra}/HETG, from 2008 November
6 -- 30,  and for 350 ksec with {\it Suzaku}, on 2008 November 23. The analysis of
these data are discussed in detail in Miller et al. (2010), Turner et al. (2010)
and Lobban et al. (2011; hereafter L2011). To summarize the results from L2011,
based on photoionization modeling and spectral analysis, we found evidence for 5
zones of absorption, spanning a range of $-180$ km s$^{-1} \leq$ $v_{r}$ $\leq$
$-710$ km s$^{-1}$ and a highly ionized zone at $v_{r}$ $\sim$ $-5800$ km
s$^{-1}$, similar to the results from S2009. Additionally, the continuum-fitting
required another zone, with a covering factor of 30\%. In this paper, we present
our analysis of {\it HST} COS observations of NGC 4051. In addition to the
characterization of the UV absorption, via spectral analysis and photoionization
modeling, we discuss the relation between the UV absorbers and the components of
X-ray absorption, the latter as modeled by L2011 and S2009.

\section{Observations and Analysis}

We obtained {\it HST} COS observations of the nucleus of NGC~4051 on 2009
December 11 (UT) as part of a multiwavelength effort. Due to scheduling
constraints, the UV observations were not simultaneous with our {\it Suzaku} and
{\it Chandra} X-ray observations. We obtained optical spectra of NGC~4051
through a 2$''$ wide slit in photometric conditions on 2009 December 17 and 18,
close in time to the COS observations, with the Perkins 1.8-m telescope and
DeVeny Spectrograph at Lowell Observatory. The optical spectra span the range
4000 -- 7000 \AA\ at a spectral resolution of $\sim$3.3 \AA. We also retrieved
{\it FUSE} archival spectra from 2002 March 29, 2003 January 18, and 2003 March
19 (Kaspi et al. 2004; Dunn et al. 2008), obtained through the 30$''$ $\times$
30 $''$ aperture, and averaged these together to improve the signal-to-noise
ratios (SNRs). Finally, we made use of the {\it HST} STIS echelle E140M
observation of of NGC~4051 by Collinge et al. (2001), obtained through a
0\arcsecpoint2 $\times$ 0\arcsecpoint2 aperture.

We observed NGC~4051 with COS over 4 {\it HST} orbits with the G130M and G160M
gratings through the Primary Science Aperture, which is 2\arcsecpoint5 in
diameter. COS far-UV observations with these gratings are imaged onto two
side-by-side detectors, and thus each observation consists of two spectra
separated by a small wavelength gap (Dixon et al. 2010; Green et al. 2012). We therefore obtained
the spectra at different wavelength offsets to ensure full coverage from 1137 --
1772 \AA. We give the wavelength coverages and exposure times of the individual
spectra in Table 1.

We extracted the one-dimensional COS spectra from the STScI ``x1d'' files and
interpolated them onto a common wavelength grid. The absolute fluxes of the
individual spectra in the regions of overlap differ by $<$5\%, and we therefore
combined them without any scaling in flux. We flagged regions at the ends of
each spectrum that were not useful (and not flux-calibrated in the pipeline
processing) as well as regions containing obvious artifacts (in particular,
shadows cast by ion repeller grid wires), and then averaged fluxes in good
regions on a pixel-by-pixel basis. Individual pixels in the final averaged
spectrum contain contributions from one to four original spectra; most are from
two spectra. The resolving power of the combined COS spectrum is
$\lambda$/$\Delta\lambda$ $\approx$ 16,000, which translates to a velocity
resolution of 19 km s$^{-1}$ ($\sim$8 pixels) FWHM. Comparison with the STIS
echelle spectra of NGC 4051 indicated a slight constant offset in the central
wavelengths of the Galactic lines; we corrected for the offset by adding $+$0.07
\AA\ to the COS spectra to allow for direct comparison with the STIS
observations.

We show the full COS spectrum and identify the principal UV emission lines in
Figure 1. The SNRs per resolution element range from 5 to 10 in the continuum
regions and up to 50 at the peaks of the emission lines, about twice those in
the STIS spectrum (for the same wavelength bins) even though the COS spectrum
was obtained when NGC 4051 was in a low flux state. In Figure 2, we show the UV
continuum light curve for NGC~4051 spanning 32 yr, derived primarily from {\it
International Ultraviolet Explorer (IUE)} observations (Dunn et al. 2006). The
STIS and COS observations occurred when NGC~4051 was in ``average'' (F[1365 \AA]
$= 1.38 (\pm 0.16) \times 10^{-14}$ erg s$^{-1}$ cm$^{-2}$ \AA$^{-1}$) and
relatively low (F[1365 \AA] $= 0.83 (\pm 0.05) \times 10^{-14}$ erg s$^{-1}$
cm$^{-2}$ \AA$^{-1}$) continuum states, respectively, although it is clear from
the previous {\it IUE} observations that large amplitude variations could have
occurred in the $\sim$10 yr time interval between these latest observations.

As noted by Collinge et al. (2001), the STIS UV spectrum of NGC~4051 contains a
complex set of absorption lines from our Galaxy, the host galaxy of NGC~4051,
and clouds of gas that are intrinsic to NGC~4051 and outflowing with respect to
its nucleus.  We list their absorption components, radial velocities relative to
the emission-line redshift, and full-width at half-minima (FWHM; we use the same
acronym for absorption and emission lines) in Table 2. In Figures 3 and 4, we
show expanded versions of the COS and STIS spectra around absorption lines
spanning a wide range in ionization (N~V $\lambda\lambda$1238.821, 1242.804;
C~IV $\lambda\lambda$ 1548.202, 1550.774;Si~IV $\lambda\lambda$1393.755,
1402.770, Si~III $\lambda$1206.500; C~II $\lambda$1334.532; Si~II
$\lambda$1260.422). We also show the radial velocities of the kinematic
components of absorption (for the strongest member of the doublet) in Figure 3.
In order to be consistent with the radial velocities of Collinge et al., we use
their systemic redshift of $z = 0.002295$ for NGC~4051 based on its optical
emission lines (de Vaucouleurs et al. 1991). Using the H~I 21-cm redshift ($z
=0.002418$, de Vaucouleurs et al. 1991) would shift the radial velocities of the
absorption lines by an additional $-$37 km s$^{-1}$. The higher SNRs of the COS
spectrum are apparent in Figures 3 and 4, and the lower spectral resolution of
COS (19 km s$^{-1}$) compared to that of the STIS E140M grating (7 km s$^{-1}$)
is also clearly evident, for example in the depth of Component 1 in N~V.

We find the same absorption components (G, 1-- 9) in the COS spectrum that were
identified by Collinge et al. (2001) in the STIS spectrum. We also confirm that
Component 8, most clearly seen in N~V, is shifted to more negative radial
velocities in lower ionization lines (Collinge et al. suggest that this is a new
low-ionization component and identify it as Component 10). We did not find any
UV absorption line corresponding to the $-$2340 km s$^{-1}$ absorption system
identified in X-ray observations (Collinge et al. 2001), in agreement with the
lack of detection of this system in the STIS (Collinge et al.) and {\it FUSE}
(Kaspi et al. 2004) spectra.

In Figure 4, the Galactic components (labeled ``G'') in the strong
low-ionization lines are completely saturated and black (i.e., consistent with
zero flux in their cores), but these same lines in the COS spectra have a slight
pedestal at $\sim$6 $\times$ 10$^{-16}$ erg s$^{-1}$ cm$^{-2}$ \AA$^{-1}$ (about
7\% of the continuum flux). The latter is likely due to instrumental scattering
by the broad wings of the COS line-spread function (Dixon et al. 2010), and we
take this extra emission into account when analyzing the absorption components.

\section{Observational Results}

We first identify those absorption components that are likely associated with
the interstellar medium (ISM) in our Galaxy or the host galaxy of NGC~4051,
based primarily on their radial velocities. Taking a closer look at Figures 3
and 4, we confirm that the ``G'' component identified by Collinge et al. (2001)
arises in our Galaxy's ISM; the absorption lines are broad, completely saturated
at low ionization, and at a heliocentric radial velocity of only $-$39 km
s$^{-1}$. Strong Galactic lines at this velocity can also be seen in the {\it
FUSE} spectrum of NGC~4051 (Kaspi et al. 2004); these lines, and in particular
the Galactic H$_2$ absorption, are studied in detail in Wakker (2006). Component
1 is present in N~V, C~IV, Si~IV and possibly Si~III, and is at a heliocentric
radial velocity of only $+$41 km s$^{-1}$; it is likely a high-ionization
component of gas in our Galaxy (see Bowen et al. 2008; Fox et al. 2006),
although an origin in NGC 4051 cannot be ruled out.
Component 8 shows broad, saturated absorption in C~IV, Si~IV, Si~III, C~II, and
Si~II at a central radial velocity of $-$48 to $-$80 km s$^{-1}$ with respect to
NGC~4051 (Collinge et al. 2001); it is likely due to gas in the disk or halo of
NGC~4051. Component 9 shows moderate absorption in these lines as well, at a
radial velocity of $+$30 km s$^{-1}$. This component and two others seen only in
Ly$\alpha$ at $+$110 km s$^{-1}$ and $+$260 km s$^{-1}$ (Collinge et al. 2001)
are likely associated with high-velocity clouds in the host galaxy of NGC~4051,
similar to those seen in our Galaxy (Fox et al. 2006; Shull et al. 2009). The remaining components
are likely intrinsic to and outflowing from the active nucleus in NGC~4051.

The troughs of the low-ionization lines of Component 8 in Figure 4 are
heavily saturated and go down to zero flux in the STIS data, but show
excess emission in the COS data. The excess flux cannot be due to
instrumentally scattered light, because it does not appear in the troughs
of the nearby Galactic lines, and there is no extra line emission near the
lines of Component 8, at least in the regions of Si~III and Si~II
$\lambda$1260. The Component 8 lines bottom out at $\sim$1.2
$\times$10$^{-15}$ erg s$^{-1}$ cm$^{-2}$ \AA$^{-1}$ in the COS data,
rather than $\sim$6 $\times$10$^{-16}$ erg s$^{-1}$ cm$^{-2}$ \AA$^{-1}$
for the Galactic lines in the same areas of the spectrum. Thus, in addition
to the instrumental scattered light, the COS spectrum show a slight excess
UV continuum at the level of $\sim$6 $\times$10$^{-16}$ erg s$^{-1}$
cm$^{-2}$ \AA$^{-1}$ in its 2\arcsecpoint5 aperture compared to the
0\arcsecpoint2 $\times$ 0\arcsecpoint2 aperture of STIS.

Intrinsic absorption components 2 -- 7 are present in N~V and C~IV, but none are
clearly evident in Si~IV or lower ionization lines, indicating that the gas in
these components is relatively highly ionized. Components 2, 5, and 7 show the
strongest lines, but Component 7 is strongly blended with Component 8 in C~IV.
The remaining components (3, 4, and 6) are relatively weak and difficult to
separate from the major components due to blending. We therefore concentrate on
Components 2, 5, and 7, because they are strong, less affected by blending, and
likely dominate the mass outflow in the UV. We also investigate the nature of
components 8 and 9 in more detail.

Convolving the STIS spectrum with the COS line-spread function (LSF, Dixon et 
al. 2010) and scaling it down to match the COS fluxes, we find no clear 
evidence for intrinsic changes in the absorption components over the $\sim$10 
yr interval between observations, as demonstrated for C~IV in Figure 5.
The apparent difference in depths of some of the absorption components
is likely due to excess emission in the COS spectra, as discussed in more detail below.
This
is somewhat unusual, because most Seyfert 1 galaxies show strong absorption
variability, including the appearance and disappearance of absorption
components, on time scales of years (e.g., NGC~3516, Kraemer et al. 2002;
NGC~3783, Gabel et al. 2005; NGC~4151, Kraemer et al. 2006; NGC~7469, Scott et
al. 2005; Mrk 279, Scott et al. 2009; NGC~5548, Crenshaw et al. 2009; and
references therein). However, this may not be too surprising, because many of
the absorption lines in NGC~4051 are completely saturated (Collinge et al. 2001;
this paper), and would not show detectable variations unless the ionic column
densities decreased dramatically or a significant fraction of a background
emission source (e.g., the broad line region) was covered or uncovered (e.g.,
see Crenshaw et al. 2004) over the 10-year interval between STIS and COS
observations. Thus, we cannot rule out ionic column density changes in the
saturated lines between the STIS and COS observations.

To further investigate the nature of the uncovered emission in the COS spectrum
and its effect on measurements of the ionic column densities, we generated a
detailed model of the background emission spectrum. We fitted a spline to the
continuum across the entire spectrum using regions unaffected by emission or
absorption features. To generate template profiles for the emission lines, we
used the He~II $\lambda$1640 line, which is not affected by absorption.
Inflections in the He~II profile, shown in Figure 6, suggest the presence of 3
components: broad, intermediate, and narrow. We fitted the broad component with
a cubic spline and the other two components with Gaussians. The resulting widths
of the emission-line components are 260, 1090, and 4500 km s$^{-1}$ (FWHM). We
associate the first and last of these with the narrow-line region (NLR) and
broad-line region (BLR), and suggest that the middle component arises in an
``intermediate-line region'' (ILR), which we have also identified in low-flux UV
spectra of NGC 4151 (Crenshaw \& Kraemer 2007) and NGC~5548 (Crenshaw et al.
2009). However, we note that our ``ILR'' likely corresponds to the ``BLR''
detected in optical spectra, which show a ``broad'' component of H$\beta$ with
FWHM $=$ 1070 km s$^{-1}$ (Peterson et al. 2004).

To model the emission-line profiles of the high-ionization lines, we reproduced
the He~II templates at the expected positions of the lines, retaining the same
velocity widths, and scaled them in intensity to obtain the best fits to the
observed profiles.. Allowing the doublet ratios (e.g., N~V
$\lambda$1238.8/$\lambda$1242.8) to vary between 1 and 2 (see Crenshaw et al.
2009) made little difference to the overall fits, so we fixed them to a value of
1. We adopted the minimum narrow-line fluxes needed to fit the observed
profiles. Figure 7 shows that our procedure yields an excellent fit to the
observed profiles in the COS spectra, and provides an accurate deconvolution of
the NLR, ILR, and BLR contributions to the emission-line profiles.

As noted above, absorption components 2, 5, and 7 are saturated at zero
flux in their cores in the STIS spectra. Their non-zero fluxes in the COS
spectra must be due to uncovered emission. To model the uncovered emission,
we use our derived values for the instrumentally-scattered light and the
excess emission from the host galaxy, plus contributions from the NLR
fluxes derived above, as we have done in the past (Kraemer et al. 2002;
Crenshaw et al. 2009). Thus, a reasonable model for the uncovered emission
is a combination of the excess continuum flux at a level of 1.2
$\times$10$^{-15}$ erg s$^{-1}$ cm$^{-2}$ \AA$^{-1}$ (half due to
instrumental light, half to extra UV flux in the COS aperture) plus the NLR
emission. As shown in Figure 6, the model provides provides a reasonably
good fit to the uncovered emission, in that it skims just beneath the
unblended saturated lines associated with Components 2, 5, and 7.
Furthermore, it demonstrates that Component 5 and 7 must be mostly inside
the NLR, whereas Component 8 absorbs most, if not all, of the NLR emission
(we cannot constrain Component 2 in this manner because it does not overlap
the NLR emission in velocity space). On the other hand, Components 2, 5 and
7 absorb most, if not all, of the continuum, BLR, and ILR emission, and
therefore lie outside of these regions.

We measured the ionic column densities of Components 2, 5, 7, 8, and 9 by
subtracting the uncovered emission components from the spectra and dividing by
the covered emission components (covered continuum, BLR, and ILR fluxes) to
obtain normalized profiles. We then converted these to optical depth as a
function of wavelength, and integrated across the optical depth profiles to
obtain column densities (Crenshaw, Kraemer, \& George 2003). 
Uncertainties in the measured column densities come from propagation of photon
noise and different reasonable fits to the underlying emission. Measurements of
unsaturated lines in both COS and STIS spectra resulted in differences
smaller than the quoted uncertainties, reinforcing our claim that these
absorption features did not vary.
Components 2, 5,
and 7 are completely saturated in N~V and C~IV, and we can only provide lower
limits to their ionic column densities. We used the higher SNR COS spectra, and
assumed the least amount of uncovered emission, given the uncertainties, to
determine these limits. To obtain upper limits to the Si~IV columns, we added
model absorption-line profiles to the spectra and decreased their equivalent
widths until they were not detectable above the noise.

We used the {\it FUSE} spectra of NGC~4051 to obtain further constraints on the
ionic column densities. The COS and {\it FUSE} observations are not concurrent,
and our underlying assumption is that the limits obtained from the COS
observations apply to the epochs of {\it FUSE} observations as well. By matching
the COS and average {\it FUSE} spectra in the region of wavelength overlap, we
find the continuum flux at 1365 \AA\ is $1.06 \times 10^{-14}$ erg s$^{-1}$
cm$^{-2}$ \AA$^{-1}$ projected from the {\it FUSE} spectra, intermediate between
the STIS and COS observations, providing further justification for the above
assumption. As noted by Kaspi et al. (2004), the O~VI absorption components are
blended together into big troughs, similar to Ly$\alpha$, and no useful limits
can be obtained. We agree with Kaspi et al. that N~III $\lambda$989.790
absorption is undetectable, primarily due to contamination by geocoronal
emission and Galactic H$_2$ absorption, and that C~III $\lambda$977.020
absorption is present in Component 5 and possibly present in Component 2
(Component 7, if present, is blended with 8). We see no evidence for P~V
$\lambda\lambda$1117.98, 1128.01 absorption, which would have indicated very
high columns of high-ionization gas, nor C~III$^{*}$ $\lambda$1175 
absorption in either the {\it FUSE} or {\it COS} spectra. As noted by Kaspi et
al., H~I Lyman absorption lines are seen extending down to the Lyman edge. We
identified H~I absorption to Ly6 $\lambda$930.748 in Component 2 and Ly9
$\lambda$920.963 in Component 5, whereas Lyman absorption in Component 7 is
always blended with that in Component 8. The longer wavelength lines of
Ly$\beta$ and Ly$\gamma$ are strongly affected by geocoronal emission and likely
saturated in any event, so we used the shorter wavelength lines to determine the
H~I columns for Components 2 and 5.  The C~III absorption profiles for
Components 2 and 5 (shown in Kaspi et al.
2004) appear not to be saturated, and we measured their columns
directly by integrating their optical depths across the profiles
after subtracting off the uncovered emission.

We present the ionic column densities or limits for Components 2, 5, 7, 8, and 9
in Table 3. Using a different technique for the {\it FUSE} analysis, Kaspi et
al. (2004) found that the H~I column density in Component 5 is 1.0$^{+0.6}_{-0.5}$
$\times$ 10$^{16}$ cm$^{-2}$. Our value is about half of theirs, but the error
bars still overlap. They do not give column densities for any of the other
lines. We do not detect O~I in any absorption component; its ionic column
density is therefore $<$ 1.0 $\times$ 10$^{14}$ cm$^{-2}$.

To gauge the reddening in the line of sight to the nucleus of NGC~4051, we
measured the total fluxes of the He~II $\lambda$1640 emission line in the COS
spectrum and the He~II $\lambda$4686 emission line in the concurrent optical
spectra. The lines are due to recombination and the intrinsic H~II
$\lambda$1640/$\lambda$4686 ratio should be $\sim$7.25, for an electron density n$_{e} = 10^{6}$ cm$^{-3}$ and
temperature T $= 5 \times 10^{4}$ K (Seaton 1978), which we chose to
match the combined BLR and NLR emission.
To ensure an accurate absolute flux, we scaled our averaged
optical spectrum slightly (by a factor of 1.15) so that our [O III]
$\lambda$5007 flux matched the well-established value of Peterson et al. (2000).
The resulting observed He~II $\lambda$1640/$\lambda$4686 ratio is then (14
$\times$ 10$^{-14}$ erg s$^{-1}$ cm$^{-2}$)/(4.0 $\times$ 10$^{-14}$ erg
s$^{-1}$ cm$^{-2}$) $=$ 3.5. This value yields E(B$-$V) $=$ 0.19 (only 0.01 from
our Galaxy) for a Galactic reddening curve (Savage \& Mathis 1979) and $N_H$ $=$
1.0 $\times$ 10$^{21}$ cm$^{-2}$ for a standard Galactic dust-to-gas ratio (Shull \& van
Steenberg 1985). Kaspi et al. (2004) present an observed spectral energy distribution (SED) for NGC~4051 which
has a UV spectral index of $\alpha$ = $-$2.0 (where $F_{\nu} =  K\nu^{\alpha}$)
between 1160 and 3000 \AA. Correction for extinction based on the above
reddening yields $\alpha$ = $-$1.3 and F(1365 \AA) $= 6.4 \times 10^{-14}$ erg
s$^{-1}$ cm$^{-2}$ \AA$^{-1}$ for the STIS observation used in Kaspi et al.'s SED.

We searched for and did not find any H$_2$ absorption in the {\it FUSE} spectrum near
the redshift of NGC~4051. Using the methods of Dunn et al. (2007), we find that
N(H$_2$) $<$ 5.0 $\times$ 10$^{14}$ cm$^{-2}$ for any rovibrational level.
Thus, the reddening is not due to dust in cold molecular gas.

\section{Photoionization Models}

The  photoionization models used for this study were generated using the
photoionization code Cloudy, version 08.00 (last described by Ferland et al.
1998). We assumed an open, or ``slab'', geometry. As per convention, the models
are parametrized in terms of the dimensionless ionization parameter, $U =
Q/(4\pi r^{2}{\rm c} n_{\rm H})$, where $r$ is the radial distance of the
absorber, $n_{\rm H}$ is hydrogen number density, in units of cm$^{-3}$ and $Q =
\int_{13.6 eV}^{\infty}(L_{\nu}/h\nu)~d\nu$, or the number of ionizing photons
s$^{-1}$ emitted by a source of luminosity $L_{\nu}$, and the total hydrogen
column density, $N_{\rm H}$ (in units of cm$^{-2}$), where $N_{\rm H} = N_{\rm
HI} + N_{\rm HII}$.

\subsection{Model Inputs}

In our previous paper (L2011), we characterized the intrinsic X-ray continuum in
the {\it Suzaku} XIS $+$ HXD (0.5 -- 10 keV and 20 -- 50 keV, respectively) as a
power-law with a photon index $\Gamma \approx 2.5$. This is consistent with
constraints on the EUV-soft X-ray SED obtained via photoionization models and
the ratios of [Ne~V] 14.32 $\mu$m, 24.32 $\mu$m, [Ne~III] 15.56 $\mu$m, and
[O~IV] 25.89 $\mu$m observed in {\it Spitzer} IRS spectra (Mel\'endez et al.
2011). Noting that the UV observations found NGC~4051 in a relatively low-flux
state, we used the lowest of the 3 {\it Suzaku} fluxes, 3.4 $\times$ 10$^{-12}$
ergs cm$^{-2}$ s$^{-1}$ at 2 keV, corrected for absorption, and extrapolated
down to 1365 \AA, assuming a spectral index $\alpha = 1.5$. This predicts a flux
of 7.4 $\times$ 10$^{-14}$ ergs cm$^{-2}$ s$^{-1}$ \AA$^{-1}$, which is close
to the dereddened flux from the STIS spectrum used by Kaspi et al. (2004).
Based on this, and the fit to the UV continuum (see Section 3), we have
parametrized the SED in the form of a broken power law, such that $L_{\nu}
\propto \nu^{\alpha}$ as follows: $\alpha = -1.3$ for energies $<$ 9.8 eV (1365
\AA), $\alpha = -1.5$ over the range 9.8 eV $\leq$ h$\nu$ $<$ 50 keV\footnote{In
L2011, the photoionization models were parametrized in terms of $\xi =
(\int_{13.6 eV}^{13.6 keV} L_{\nu}~d\nu)/(n_{\rm H} r^{2})$; for our assumed
SED, $U \approx 0.04 \times \xi$.}.  We included a low energy cut-off at $0.1$ eV 
and a high energy cutoff at 100 keV. Using this SED
and the unabsorbed flux at 2 keV, we determined that $Q \approx 6.4 \times
10^{52}$ photons s$^{-1}$. For these models, we have assumed roughly solar
elemental abundances (e.g., Asplund et al. 2009) as follows (in logarithm,
relative to H, by number):
He: $-1.00$, C: $-3.57$, N: $-4.17$, O: $-3.31$, Ne: $-4.07$,
Na; $-5.76$, Mg: $-4.40$,  Al: $-5.55$, 
Si: $-4.49$,  P: $-6.59$, S: $-4.88$,  Ar: $-5.60$, 
Ca: $-5.66$,  Fe: $-4.50$, and Ni: $-5.78$. 

In these spectra, we were only able to derive column densities for a small
number of ions in the various kinematic components (see Section 3 and Table 3).
Therefore, our modeling method is to fit these derived columns, within the
measurement uncertainties, exceed lower limits for the saturated lines, and
under-predict upper limits for those ions which were not detected. To achieve
this, we used the ``optimize'' command in Cloudy, adjusting the commandable
errors as needed to improve the fit (see Cloudy Manual, HAZY; Ferland 1996).
However, uncertainties in both elemental abundances and atomic data (e.g.,
dielectronic recombination rates) can affect the predictions of ionic column
densities, particularly for poorly populated ionic states. The final model
parameters for the strongest absorption components and the predicted ionic
column densities are listed in Tables 4 and 5, respectively.

\subsection{Model Results}

As discussed in Section 2, components 2 -- 7 are associated with mass outflow,
while components 8 and 9 are more likely formed in the ISM or halo of the host
galaxy. Of the mass-outflow components detected in the STIS and COS spectra, the
strongest are components 2, 5, and 7. For components 2 and 5, we have measured
column densities for H~I and C~III (see Table 3), lower limits for N~V and C~IV,
and an upper limit for Si~IV. 
Given these limited constraints, models spanning a range of $U$ and $N_{\rm H}$
can fit the measured values. For example, using the measured columns for H~I and C~III and
their respective uncertainties,  and the lower limit for C~IV, acceptable models parameters for component 5 are
log$U= -0.45^{-0.2}_{+0.1}$ and log$N_{\rm H} = 20.74^{-0.40}_{+0.25}$. However, for component 5,
$v_{r} = 268$ km s$^{-1}$ and
FWHM $=$ 133 km s$^{-1}$, which are quite close to the values determined by S2009 for their
zone 4, i.e., $v_{r} = 270$ km s$^{-1}$ and FWHM $=$ 170 km s$^{-1}$. Furthermore, the S2009 zone 4 model
parameters, log$U= -0.64$ and log$N_{\rm H} = 20.4$, are within the range acceptable
for component 5, hence we have opted to use these values. The predictions for this model are a 
good fit for component 5, albeit with a slight overprediction of the C~III
column density. To summarize, while our final model parameters could be off by 
0.2dex in $U$ and 0.4dex in $N_{\rm H}$, requiring consistency with the X-ray results can, at least,
direct us to a preferred location in $U$-$N_{\rm H}$-space (we discuss the relationship between the UV and X-ray absorption in
detail in Section 5). 

For component 2, there are no clear X-ray constraints and our best-fit model overpredicts the H~I column
density and slightly underpredicts the lower limits for N~V and C~IV. Na\"ively,
one might assume that this is an indication of super-solar N/H and C/H ratios.
Although that cannot be ruled out, increasing the carbon abundance results in a
factor of $\gtrsim$ 2 overprediction of C~III. The physical parameters of
component 7 are even less well-constrained by the observations, because only lower
limits for C~IV and N~V and an upper limit for Si~IV can be determined, and
other ions, including H~I, are blended with component 8. However, we are able to
fit the available constraints using model parameters adapted from zone 2 (L2) in
L2011.

As shown in Figures 3 and 4, while there is N~V and C~IV absorption spanning the
velocities of components 7 and 8, the lower ionization lines, e.g., C~II, Si~III
and Si~II, appear to be associated primarily with component 8. This is
consistent with our parametrization for component 7, since our model predicts
column densities for these ions $N_{\rm ion}$ $<$  10$^{12}$ cm$^{-2}$. In
Figure 3, the C~IV line is saturated for component 8, N~V is weak, and Si~IV is
strong, but not saturated. On the other hand, C~II, Si~III, and, possibly, Si~II
appear saturated, and at a more positive radial velocity than the
high-ionization lines, which is why there were identified as a separate
component (10) in Collinge et al. (2001). This suggests that there are two
physical sub-components near this velocity. One other constraint is that there
is no detectable C~II$^{*}$, which we find requires that $N_{\rm CII*}$ $<$ 1.3
$\times$ 10$^{13}$ cm$^{-2}$. Using the ratio of the upper limit to the column
density for C~II$^{*}$ to the lower limit for C~II (see Table 4), we obtain an
electron density $n_{e} \approx 1 $ cm$^{-3}$, for an electron temperature T$_{e} =
8 \times 
10^{3}$ K (e.g., Srianand \& Petitjean 2000); the density must be somewhat
lower than this value, since it was estimated using the lower limit for
$N_{\rm CII}$,. For the lower ionization model, 
8L, with $n_{\rm H} = 0.36 cm^{-3}$, logU $=$ $-3.5$ and log$N_{\rm H} = 18.8$, we
were able to match the $N_{\rm CII*}$ constraint and obtain acceptable
predictions for the column densities or lower limits for C~II, Si~III, Si~II,
Fe~II. For the higher ionization component, 8H, we were able to match $N_{\rm
NV}$, within the uncertainties, and obtained a reasonable fit for $N_{\rm SiIV}$
with logU $=$ $-2.4$ and log$N_{\rm H} = 19$, which requires that $n_{\rm H}$=
0.03 cm$^{-3}$, if the two sub-components are co-located. Based on the model parameters
and our estimate of $Q$, component 8 is $>$ 12.5 kpc from the AGN, hence,
possibly in the halo of the host galaxy. Note, this is a larger radial distance
than that found for the similar low-density, low-ionization components, A and C,
in NGC 4151, which were determined to be at radial distances of 681 pc and 2.15
kpc, respectively (Kraemer et al. 2001). Finally, component 9 shows only weak
C~IV and lower-ionization ions (Table 3). We obtained a reasonable fit, save the
over-prediction of Si~III, which could be due to uncertainties in third-row
element dielectronic recombination rates (e.g., Ali et al. 1991). As with
component 8, the absence of CII$^{*}$ suggests a low density. Assuming the same
$n_{H}$ as for 8L, component 9 would be 19.3 kpc from the AGN.

Our models for components 2, 5, and 7 predict $N_{\rm OVI} > 10^{16}$ cm$^{-2}$
(see Table 5). To
test whether the predictions are consistent with the data, we plotted the
normalized flux for O~VI as a function of velocity, using the measured values of
FWHM and $v_{r}$ for each of the three components. As shown in Figure 8, the
combined profile is characterized by a deep, square trough, spanning velocities
$-$700 km s$^{-1}$  $\lesssim  v \lesssim 0$ km s$^{-1}$, in good agreement with the 
{\it FUSE} O~VI $\lambda$1032 profile
presented in Kaspi et al. (2004; see their Figure 3), except for the 
presence of some uncovered emission in the latter. However, while our UV model
predictions are consistent with the observed O~IV absorption, we cannot rule out
the presence of additional absorption over this range in velocity, as may be
present in more highly ionized gas (i.e., the X-ray absorbers).

None of the component models predict $N_{\rm H}$ $>$ 2.5 $\times 10^{20}$
cm$^{-2}$. Coupled with their relatively high values of $U$, one consequence of
the low column densities is that none of the three strongest components is
sufficiently optically thick to the ionization radiation to justify including
the effects of screening by intervening absorbers, e.g. as is the case in NGC
3516 (Kraemer et al. 2002; Turner et al. 2005) and NGC 4151 (Kraemer et al.
2001; 2005; 2006). Also, while, as noted in Section 3, the continuum and BLR
emission is reddened, consistent with an intervening column of $N_{\rm H}$ $=$
1.0 $\times$ 10$^{21}$ cm$^{-2}$ of dusty gas, none of the models for the UV
absorber components has a sufficient column density to account for the
reddening. In fact, the sum of the column densities of components 2, 5, 7, 8,
and 9 is only 5.4 $\times$ 10$^{20}$ cm$^{-2}$.  Furthermore, the presence of
Fe~II in component 8, along with the likelihood, confirmed by our photoionization model, that most
of the iron is in the form of Fe~III, hence not detectable in these spectra, 
indicates that there is very little depltion of iron onto dust grains. Hence, there
cannot be a significant amount of dust in this component. One possibility is
that the dust is distributed within components 2, 5, and 7, and that additional
dust exists in the higher ionization gas. However, there are no tight
constraints from the modeling of the UV or X-ray absorption (e.g. L2011) that
can provide a definitive answer, hence we must leave the question as to origin
of the reddening open.

\section{Physical Properties of the Absorbers}

\subsection{Connection with the X-ray absorbers}

As shown in Table 5, the models for UV components 2, 5, and 7 predict $N_{\rm OVII}$
$>$ several $\times$ 10$^{16}$ cm$^{-2}$. Hence, these components may be
detectable in the HETG spectra.  On the other hand, the UV component models all
predict $N_{\rm OVIII}$ $<$ 10$^{16}$ cm$^{-2}$, which suggests the presence of more
highly ionized gas without a strong UV ``footprint''. In order to compare the
X-ray and UV absorption in detail, we first compared the kinematic profiles of
strong absorption lines, specifically O~VII and O~VIII with N~V and C~IV.
Initially we found no overlap of velocities, because the X-ray absorption lines
are offset to more negative velocities than the UV. However, this velocity
offset is likely due to the calibration of the {\it Chandra} gratings (see the
{\it Chandra} Proposer Guide:
http://cxc.harvard.edu/proposer/POG/html/HETG.html). For example, in L2011, we
had used only the MEG in the 0.5-1.0 keV regime. The accuracy of the wavelength
scale for MEG is $\pm$0.011\AA. Hence, for O~VII, this yields absolute
uncertainty in velocity of $\pm$150 km s$^{-1}$, while for Ne~IX it would be
$\pm$ 235 km s$^{-1}$. For Si~XIII, we used  the MEG and HEG combined. For the
HEG, the accuracy of the wavelength scale is $\pm$0.006 \AA~, which results in
an uncertainty for Si~XIII of $\pm$ 250 km s$^{-1}$. This justifies some
flexibility in comparing the X-ray and UV line kinematics. Indeed, after we
applied a $+$ 200 km s$^{-1}$ shift to the X-ray spectra, we found visual
correlation between the UV and X-ray absorption, as shown in Figure 9. The
trough of the O~VII profile encompasses UV components 2 through 8, and, while
the deepest part of O~VIII profile covers UV components 1 though 5, there is
significant O~VIII absorption blueward of component 1.

In L2011, we determined there were 6 separate zones of absorption. The highest
velocity zone, at $v_{r}$ $\sim$ -0.02c, in which H and He-like Fe lines form, is
too highly ionized to produce detectable UV lines and, in fact, no UV absorption
is detected at this velocity. And, as we have found in previous studies (e.g.
Turner et al. 2005), no lines are detected from the high-ionization,
high-column, partial-covering zone. For the remaining zones, we list the model
parameters, with $U$ given for the equivalent $\xi$, and possible associated UV
components, in Table 6.

The two lower ionization X-ray absorbers, zones L1 and
L2, should produce strong UV signatures. With our velocity correction to the
{\it Chandra} spectra, L1 and L2 should be near the systemic velocity, which
would associate them, kinematically, with UV components 7, 8, and 9. As we noted
above, a model using $U$ and $N_{\rm H}$ corresponding to zone 2 provides a good
match to UV component 7, which suggests that these indeed are the same
component. Given that, one would expect that L1 is associated with UV components
8 and 9. However, the column density determined by L2011 for L1 is $\sim$ 30
times larger than that of our models 8L, 8H and 9 combined. Furthermore, using
the parameters for L1 (log$U = -2.25$, log$N_{\rm H} = 20.4$), a
Cloudy-generated model predicts heavily saturated Si~IV, with log $N_{\rm SiIV}
> 15$, of which there is no evidence in the STIS and COS spectra. Therefore,
while we find significant absorption near systemic, in agreement with the HETG
analysis, the column density derived from the UV analysis is much smaller than
that derived from the X-ray. The source of the discrepancy may be the
uncertainty in the strength of the $\sim$0.1 keV black-body component used in
the parametrization of the HETG data. If the black-body were weaker than
assumed, the continuum fitting would require less low-ionization absorption.
For example, if we leave the black-body in the spectrum, but allow
its normalization to adjust, we find a 90\% lower limit on the column of 
zone L1 of log $N_{\rm H} > 19.47$  which would be in better agreement
with UV components 8 and 9. The other zones (L2-L4) are unaffected, 
as they are well constrained by the strong absorption lines in the spectrum.
However, although the exact form of this emission is uncertain,
removing the black-body component completely leads to an unacceptable
fit to the data.

L2011's zones L3a and L3b are significantly more ionized than the modeled UV
components (see Table 6), hence they do not predict any significant UV
absorption. However, it is possible that the absorbers are inhomogeneous, i.e.,
part of a multi-phase wind (e.g., Krolik \& Kriss 2001; Chelouche \& Netzer
2005) and each kinematic component included gas of a range of
ionization/density. As such, it is conceivable that L3a and UV component 4, L3b
and UV component 1, and L4 and UV component 2 are are associated. The weak N~V
and C~IV detected in component 1 and 4  may arise in small, dense knots within
the higher ionization gas. For component 2, the predicted $N_{\rm OVII}$ is much
smaller than for L4, which, again, suggests embedded low ionization gas. 

As noted above, S2009 used a 4 zone model for the absorbers in their analysis of the LETG
spectrum of NGC 4051 (we will refer to their zones as S1-4). As discussed in L2011,
there is rough agreement between S2009's high velocity component, S4, and the
highest velocity component in the HETG analysis. The model parameters for the
three lower-velocity zones are listed in Table 6. S3 is at roughly the same
$v_{r}$ and has similar $U$ and $N_{\rm H}$ as L3b. In terms of $U$ and $N_{\rm
H}$, the best agreement across the three data-sets is for UV component 7, L2,
and S1. However, both L2011 and S2009 claim much larger velocity dispersions than
permitted by the FWHM of the UV lines. One possibility is that a nearby weaker
UV component, in this case component 6, contributes to the X-ray absorption. To
illustrate this, in Figure 10 we show a simulation of the case of two absorber
components at the velocities of 6 and 7, assuming $\tau = 10$ in their cores.
When the profiles are smoothed to the velocity resolution of the LETG, $\sim$
300 km s$^{-1}$, we obtain a FWHM $\sim$ 250 km s$^{-1}$, similar to that
determined for S1. To further explore this, we generated a model for UV
component 6, with log$U = -0.08$ and log$N_{\rm H}$ = 20.0, which predicts
$N_{\rm NV} = 5.0 \times 10^{13}$ cm$^{-2}$, compared to the measured value of
$N_{\rm NV}$ $=$ 5.0 ($\pm$ 2.0) $\times$ 10$^{13}$ cm$^{-2}$, and $N_{\rm
OVII}$ $\sim$ several $\times$ 10$^{16}$ cm$^{-2}$, similar to L2011's model for
L2. Hence, the broad O~VII suggested by the fit for S1 and L2 is consistent with
similar contributions from UV components 6 and 7.

Although there is general agreement between the UV components and the results of
the two X-ray analyses, there are some clear discrepancies. First,  although S2009
did not require a component similar to L4, it is likely due to the fact that the LETG
has lower sensitivity at short wavelengths than the HETG, hence may not have detected
the lines from ions such as Mg~XII, Si~XIV that were fit by including L4.   
Also, while we identified S2 with UV component 5, there is no corresponding component required in the
HETG analysis. However, it is possible, given the uncertainties in velocities, that
L2 could correspond to either S1, as we suggested above, or S2. This is most 
likely evidence of the difficulty in
deconvolving relatively low-ionization components of absorption, with radial
velocities $\lesssim$ a few 100 km s$^{-1}$ of each other, solely based on the
analysis of X-ray spectra.
While S2009 did not require a component similar to L3a, and the relative constancy of the UV
absorbers is evidence
against temporal variations, since L3a has no strong UV signature, it is
possible that it was weaker during the LETG observations.
Finally, there
was no evidence for low-velocity, low-ionization gas, i.e., L1, in the LETG
spectrum.  In the latter case, the models for UV components 8 and 9 do not
predict significant X-ray absorption (see Table 5), hence it is likely that the
differences are due to the parametrization of the black-body component, as
suggested above. 

As noted in Section 1, one possible acceleration mechanism for mass outflows is
radiation pressure. Radiative driving requires that the inverse of the ratio of
the luminosity of source to its Eddington limit, $L/L_{\rm Edd}$, exceeds the
force multiplier ($FM$), or ratio of the total photo-absorption cross-section,
including bound-free and bound-bound transitions, to the Thomson cross-section
(e.g. Arav et al. 1994). For UV components 2, 5, and 7, the values of $FM$, at
the illuminated face, are 1.57 $\times$ 10$^{2}$, 1.35 $\times$ 10$^{2}$, and
1.81 $\times$ 10$^{2}$, respectively. Based on the estimated bolometric luminosity, $L_{\rm bol} = 10^{43}$ ergs s$^{-1}$,
from Vasudevan \& Fabian (2009), 
$L_{\rm bol}/L_{\rm Edd}$ $\sim$ 0.05 for NGC 4051, hence,
based on our models, each of
these components could be radiatively-driven. For comparison, X-ray zones L3a
and L3b have $FM$ values of 16.8 and 9.7, respectively, which are marginally
consistent with radiative driving, in particular if the central source
were somewhat more luminous in the past. On the other hand, L4 has a $FM = 1.9$, which
suggests another acceleration mechanism (see Section 1), unless the gas had been
accelerated to its present velocity during a period in which the source was significantly more
luminous. The difference in $FM$ for L4 and UV component 2 suggests complex
dynamical interactions, if these components are indeed co-located.

\subsection{Radial Distances and Density Constraints}

Both L2011 and S2009 derived constraints on the radial distances of the X-ray
absorbers. Based on the assumption that the physical depths of the absorbers do
not exceed their radial distances (which is a requirement for models that assume
open geometry), L2011 obtained upper limits for the radial distances for their
four lowest-ionization components components (see their Table 8). S2009 used the
lack of variability to obtain lower limits for components S1, 3, and 4 (see
their Table 5). For those components which were identified in both studies (see
Table 6), the lower limits obtained by L2011, are consistent with the upper
limits from S2009. Based on recombination time arguments, the decrease in
ionization parameter for S2,, required by their modeling of the variations over
a 20 ksec period, indicates a radial distance of $\leq$ 3 $\times$ 10$^{17}$ cm,
roughly consistent with the estimate for the lowest ionization component modeled
by Krongold et al. (2007). Overall, this suggests that the limits on radial
distances determined by S2009 are consistent with the other studies and can be
used to constrain the radial distances and densities of the UV absorbers.

In Table 4, we list the estimated radial distances and densities, based on
$Q$ and $U$, for the UV absorbers. As noted above, we have associated UV
component 5 with S2. Using the upper limit for radial distance, $r$, from S2009,
$<$ 3 $\times$ 10$^{17}$ cm, we derive an lower limit for  $n_{\rm H} < 7.9
\times 10^{6}$ cm$^{-3}$. Assuming UV component 7 and S1 are the same, for the
lower limit if $r$ $>$ 9 $\times$ 10$^{17}$ cm, we derive an lower limit for
$n_{\rm H} < 1.3 \times 10^{6}$ cm$^{-3}$  Although component 2 is much lower
ionization than the X-ray components closest in radial velocity, S3 or L4, for
the sake of comparison we may assume that it lies somewhere within the range of
radial distances estimated by S2009, e.g., $r$ $\sim$ 0.5pc. These estimates
place each of these absorbers outside the region which we have defined as the
ILR (see Section 2), but within much of the NLR, hence are consistent with the
evidence that components 2, 5, and 7 do not cover the NLR, as discussed in
Section 3. Assuming L4 is at the same radial distance as component 2, log
$n_{\rm H}$ $=$ 3.28. The predicted gas pressures at the ionized face for
component 2 and L4 are 3.41 $\times$ 10$^{-6}$ dyne cm$^{-2}$ and 2.73 $\times$
10$^{-7}$ dyne cm$^{-2}$, respectively. Therefore, these two components would
not be in full pressure equilibrium and component 2 would expand and dilute over
time,  unless confined by some other physical process, e.g. magnetic confinement
(see discussion in Gabel et al. 2005).

Based on the model parameters and distance constraints we can estimate the mass
loss rates for the UV absorbers, $\dot{M} = 8\pi r {\rm N_{\rm H}} {\rm m_{p}}
\mu {\rm v_{r}} C_{g}$, where the mean atomic mass per proton $\mu = 1.43$ for
solar abundances and $C_{g}$ is the global covering fraction of the absorbers.
We obtain the following rates: component 2, $\sim 0.01 C_{g}$ M$_\odot$
yr$^{-1}$; component 5, $< 1.76 \times 10^{-3} C_{g}$ M$_\odot$ yr$^{-1}$;
component 7, $> 1.37 \times 10^{-3} C_{g}$ M$_\odot$ yr$^{-1}$. The combined
mass loss from these components is on the same order as the fueling rate
($\approx$ 0.004 M$_\odot$ yr$^{-1}$; Turner et al. 2010). However, as suggested by L2011,
most of the mass loss is from the high ionization/high velocity component and
components L3a and L4. Greater mass loss in more highly ionized gas than in the
UV absorbers has been claimed for other AGN (e.g., NGC 5548; Crenshaw et al.
2009). This is suggestive of a scenario in which the UV absorbers are knots or
condensations in a more highly ionized wind, with the bulk of the mass contained
in the latter. 

\subsection{Relation to the Emission-line gas}

Cecil et al. (2002) suggested that some components of the NLR emission-like gas
in NGC 1068 would resemble intrinsic UV absorbers if viewed along the
line-of-sight to the AGN. Crenshaw \& Kraemer (2007) determined that emission
lines from the ILR of NGC 4151 could arise in the same gas responsible for the
UV and soft X-ray absorption. In photoionization modeling studies of the NLR in
NGC 4151 (Kraemer et al. 2000), NGC 1068 (Kraemer \& Crenshaw, 2000a; 2000b),
Mrk 3 (Collins et al. 2009), and Mrk 573 (Kraemer et al. 2009), in order to
account for the strength emission lines such as [Ne~V] $\lambda$3426 and [Fe~V]
$\lambda$6087, we argued for the presence of relatively highly ionized, e.g.
log$U \sim -1$, matter-bounded gas within 50 pc of the AGN. Although the column
densities of these components are typically several times greater than what we
have determined for components 2, 5, and 7 in NGC 4051, they span the same range
in $U$. Hence it is plausible that we are viewing the BLR/central source in NGC
4051 through gas in the inner regions of the NLR, in agreement with the
determination of our line-of-sight based on the emission-line kinematics, which
places our view at a polar angle of $\sim$12\deg\ with respect to the NLR bicone
axis and through the NLR material (Fischer et al. 2011). Interestingly, as is
the case in NGC 5548 (Crenshaw et al. 2009), we find no evidence that the ILR is
viewed in absorption, which suggests that the ILR is located at relatively
larger polar angle.

\section{Summary}

As part of a coordinated program, including {\it Chandra}/HETG  and {\it Suzaku}
observations, to study the nature of intrinsic absorption in the NLSy1 NGC 4051,
we have obtained and analyzed new COS spectra. Based on the spectral analysis
and photoionization modeling, we have determined the following:

1. We detect the same components of UV absorption identified in STIS spectra by
Collinge et al. (2001). None of the absorbers have shown clear changes in column
density or kinematic profiles, which indicates the absorption lines of H~I, N~V
and C~IV in the strongest components, 2, 5, and 7, are saturated, while the
weaker components have low densities and, hence, have not responded to changes
in the ionizing flux over the $\sim$ 10 years separating the  the STIS and COS
observations..

2. Of the UV kinematic components, 2 -- 7 (and possibly 1) are associated with
mass outflow. Based on photoionization modeling, we identified component 5 as
zone 4 from the analysis of the LETG spectrum by S2009. Component 7 is likely
the same zone 1 from S2009, and zone 2 from our analysis of the HETG spectra,
although the line widths required by the X-ray analyses indicate a by a second
component, which we suggest is component 6. Most of the UV components are
likely to be clumps in a more highly ionized, lower density wind associated
with the X-ray absorbers.

3. The predicted force multipliers for the three strongest UV components, 2, 5,
and 7, exceed the inverse of $L_{\rm bol}/L_{Edd}$ for NGC 4051, hence it is plausible
that they have been accelerated by radiation pressure. On the other hand, we
find that X-ray component L4 may be too highly ionized to be radiatively driven,
which suggests a complex interaction with UV component 2 if they are co-located.

4. The combined mass loss rate from these UV components is $\approx$ 0.01 $C_{g}$
M$_\odot$ yr$^{-1}$, which is on the same order as the estimated fueling rate.
Nevertheless, as discussed in L2011, the mass loss is dominated by the
contributions from more highly ionized absorbers, hence it is likely that the total mass 
loss rate significantly exceeds the fueling rate.  

5. As in the case of NGC 5548 (Crenshaw et al. 2009), there is no evidence that
the UV absorbers are associated with the ILR. Rather, the ionization parameters
and column densities determined for components 2, 5, and 7 are in the range of
those found for relatively high ionization gas in multi-component modeling of
the NLR in several Seyfert galaxies. This suggests that we are viewing the AGN
in NGC 4051 through the NLR gas, in agreement with the geometry derived from
kinematic modeling (Fischer et al. 2011) and the constraints on the radial
distances of the UV absorbers.

\acknowledgments

Support for program 11834 was provided by NASA through a grant from the Space
Telescope Science Institute, which is operated by the Association of 
Universities for Research in Astronomy, Inc., under NASA contract NAS 5-26555.
This research has made use of the NASA/IPAC Extragalactic Database (NED) which
is operated by the Jet Propulsion Laboratory, California Institute of
Technology, under contract with the National Aeronautics and Space
Administration. This research has made use of NASA's Astrophysics Data System
Bibliographic Services. Some/all of the data presented in this paper were
obtained from the Multimission Archive at the Space Telescope Science Institute
(MAST). Support for MAST for non-HST data is provided by the NASA Office of
Space Science via grant NNX09AF08G and by other grants and contracts. We thank
Gary Ferland for his continued maintenance of the code Cloudy.

\clearpage

\figcaption[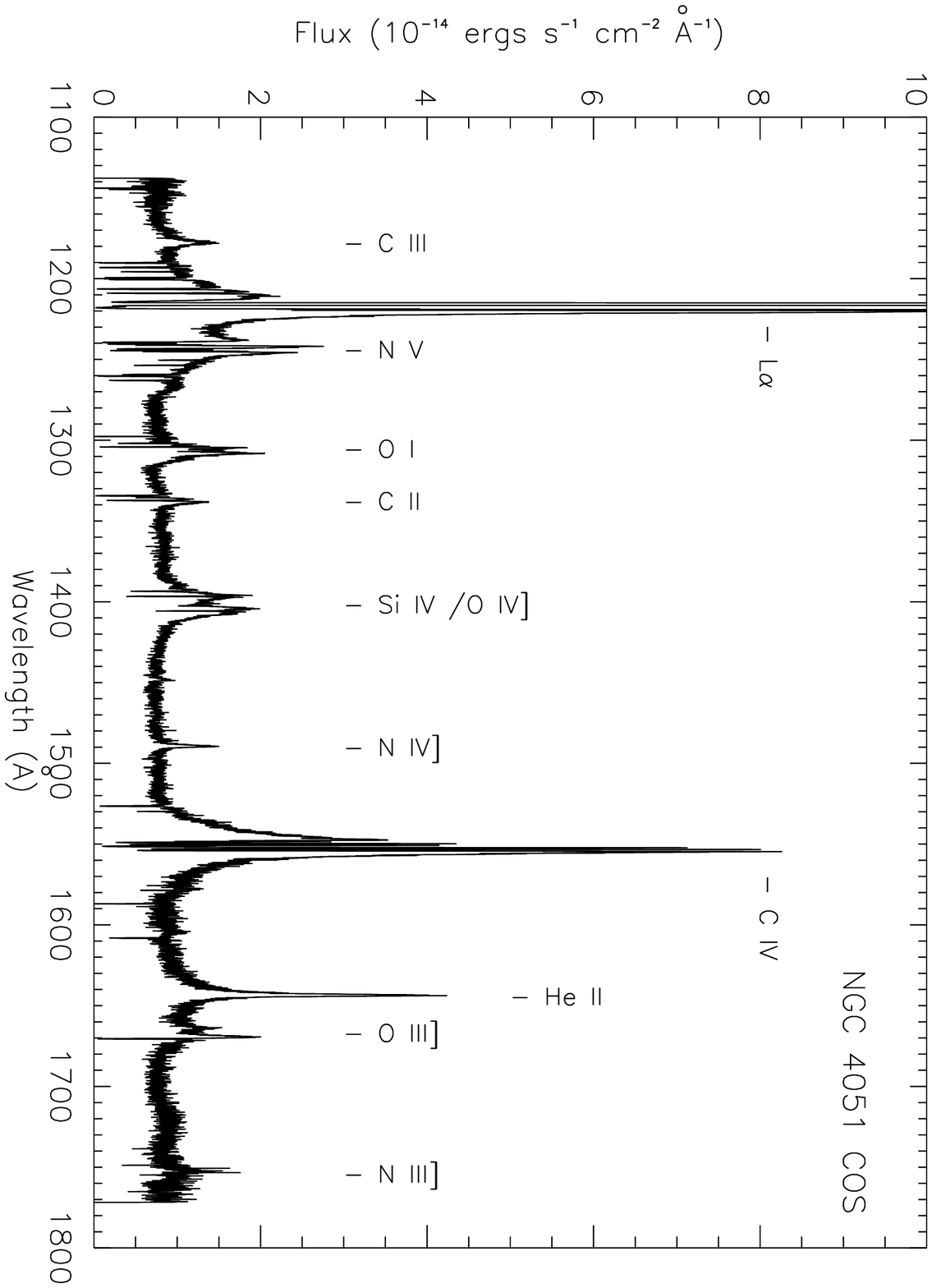]{Combined COS spectrum of NGC~4051 in the observed frame,
smoothed with a 7-point boxcar. Prominent emission lines are labeled.}

\figcaption[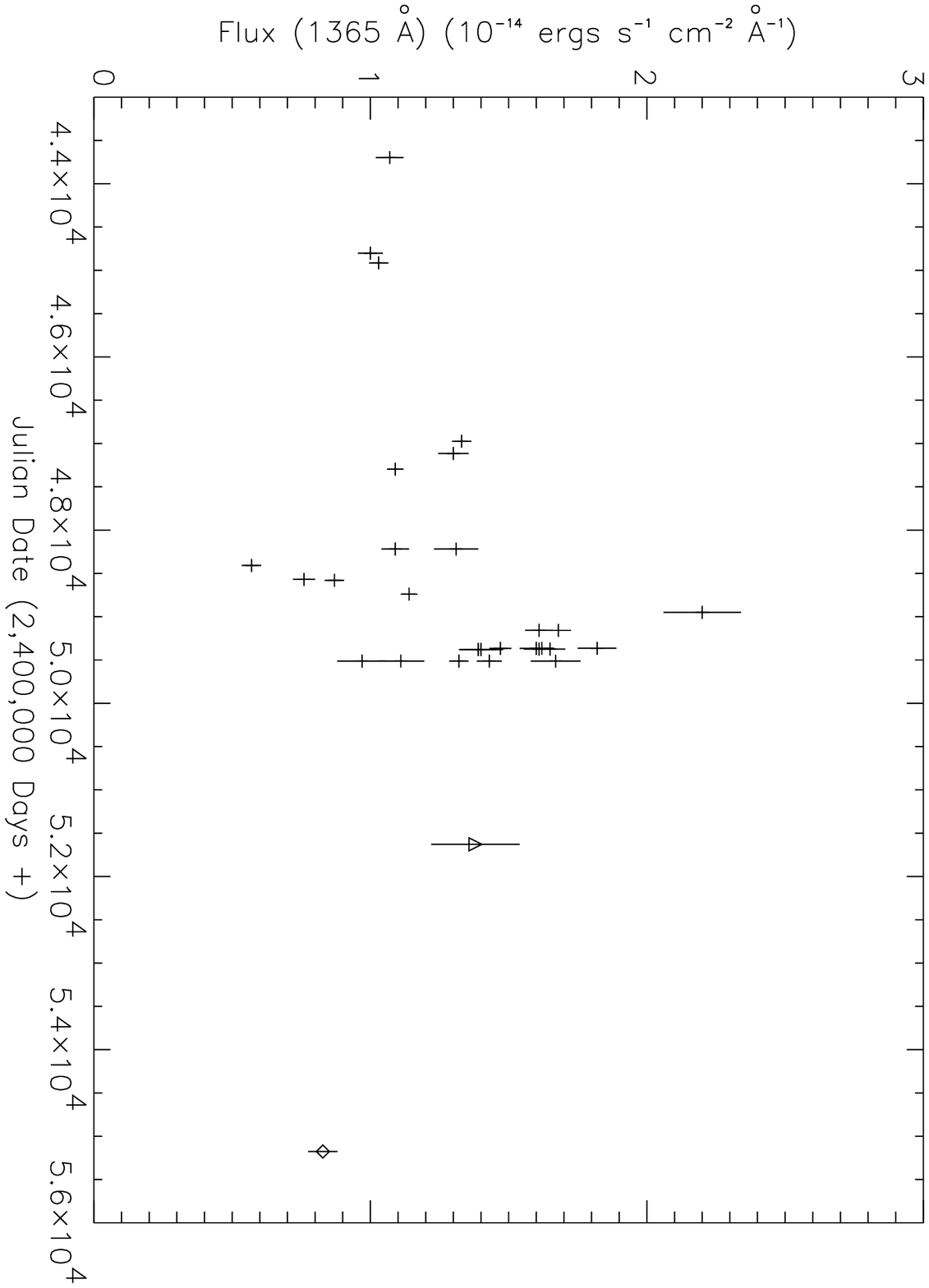]{UV continuum light curve of NGC~4051. Fluxes at 1365~\AA\ 
are plotted as a function of Julian date. The symbols are as follows: pluses --
{\it IUE}, triangle -- STIS, diamond -- COS. Vertical lines indicate the
one-sigma uncertainties.}

\figcaption[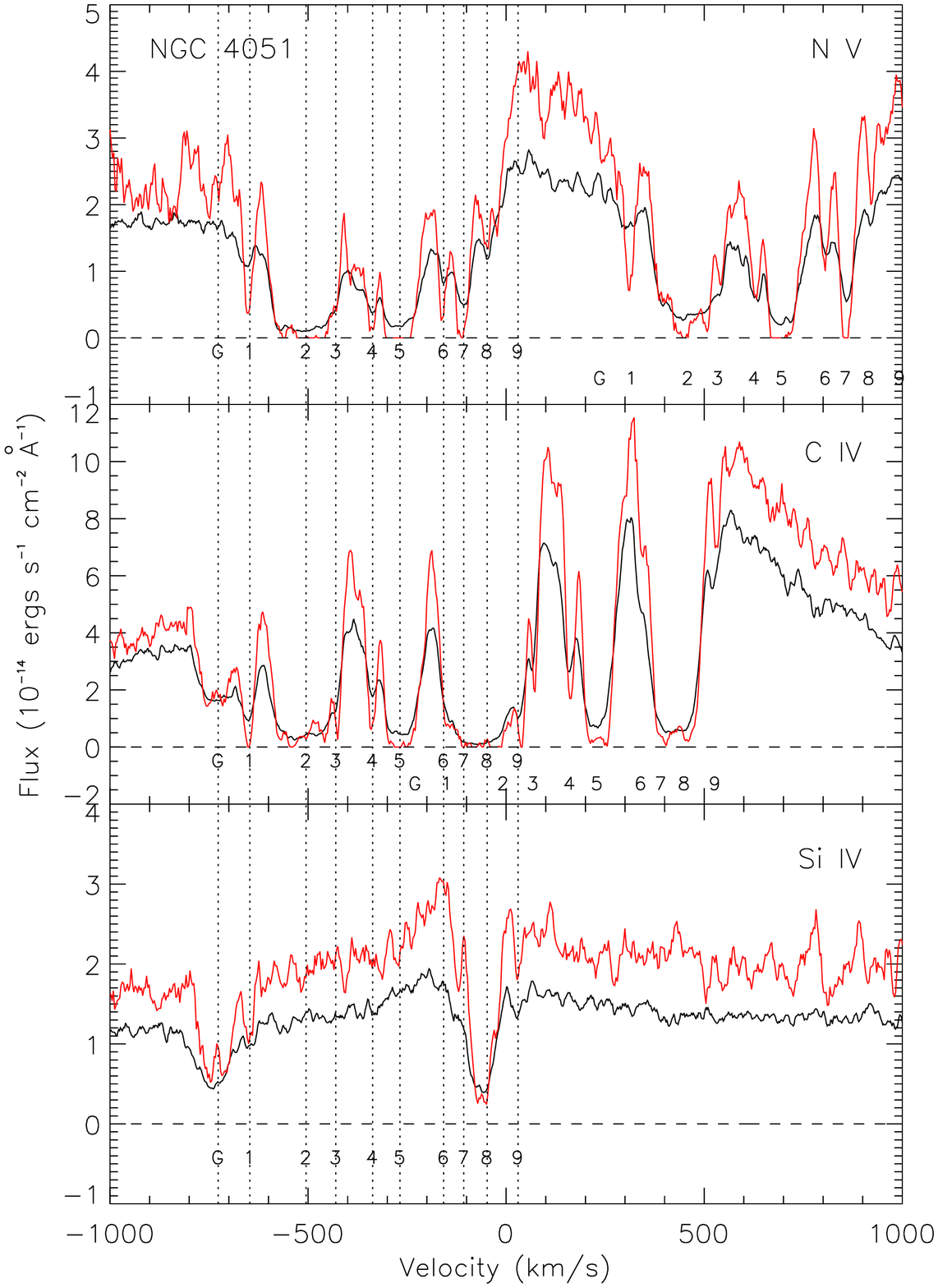]{Kinematic components of absorption for high-ionization lines
in the UV spectrum of NGC~4051, originally identified by Collinge et al. (2001).
The black line gives the COS spectrum and the red line gives the STIS spectrum.
Fluxes are plotted as a function of radial velocity for the strongest member
of each doublet, relative to an emission-line redshift of z $=$ 0.002295.
The kinematic components are identified for both members of each doublet.}

\figcaption[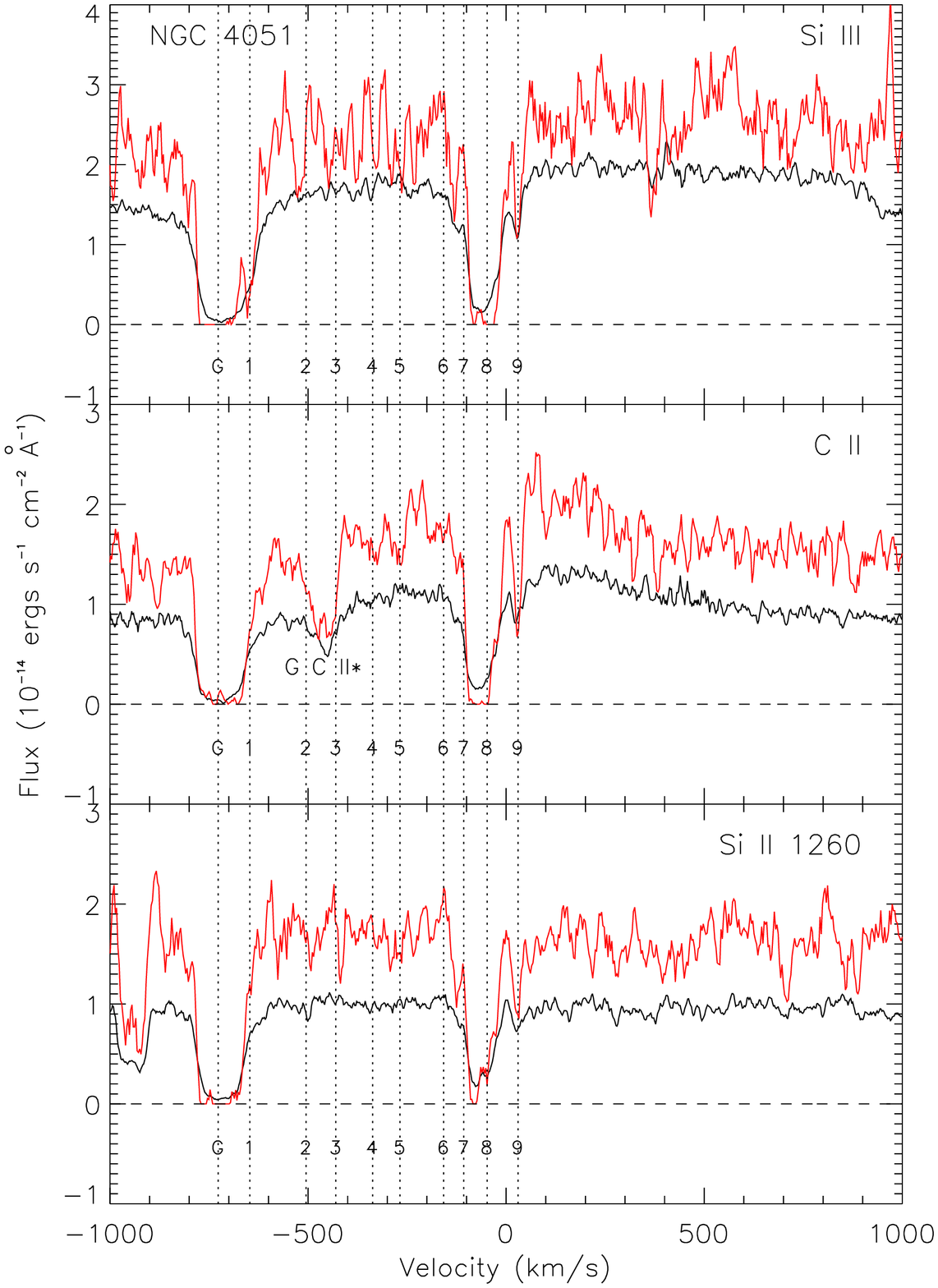]{Kinematic components of absorption for low-ionization lines
in the UV spectrum of NGC~4051. The black line gives the COS spectrum and the
red line gives the STIS spectrum. Fluxes are plotted as a function of radial
velocity for each line, relative to an emission-line redshift of z $=$
0.002295.}

\figcaption[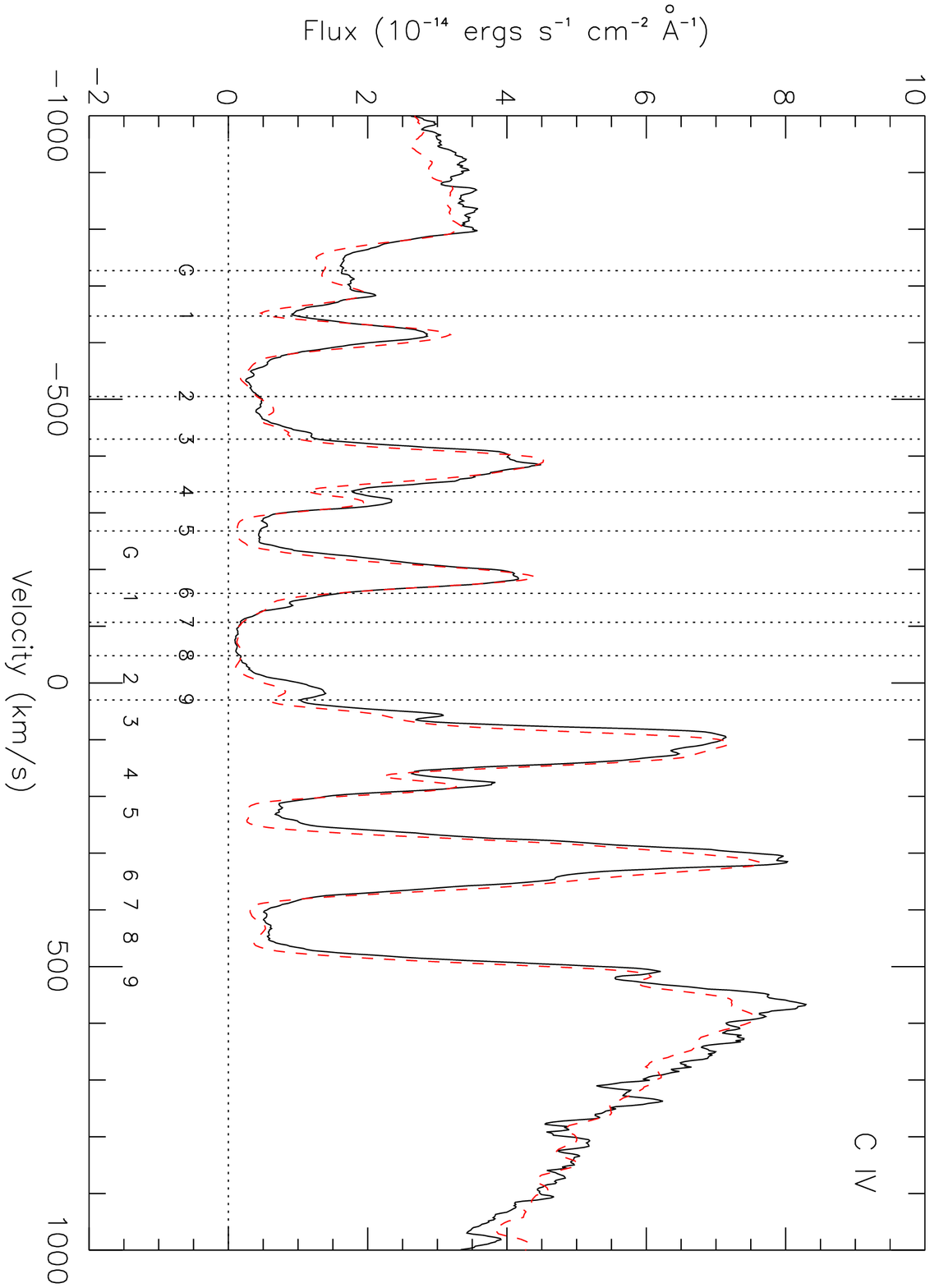]{Comparison of the C~IV profiles from the COS (black line) 
and STIS spectra (red dashed line). The latter has been convolved with the 
COS LSF and scaled by a factor of 0.75. The only clear difference is excess 
emission in some of the absorption troughs, which is due to uncovered 
emission as described in the text.}

\figcaption[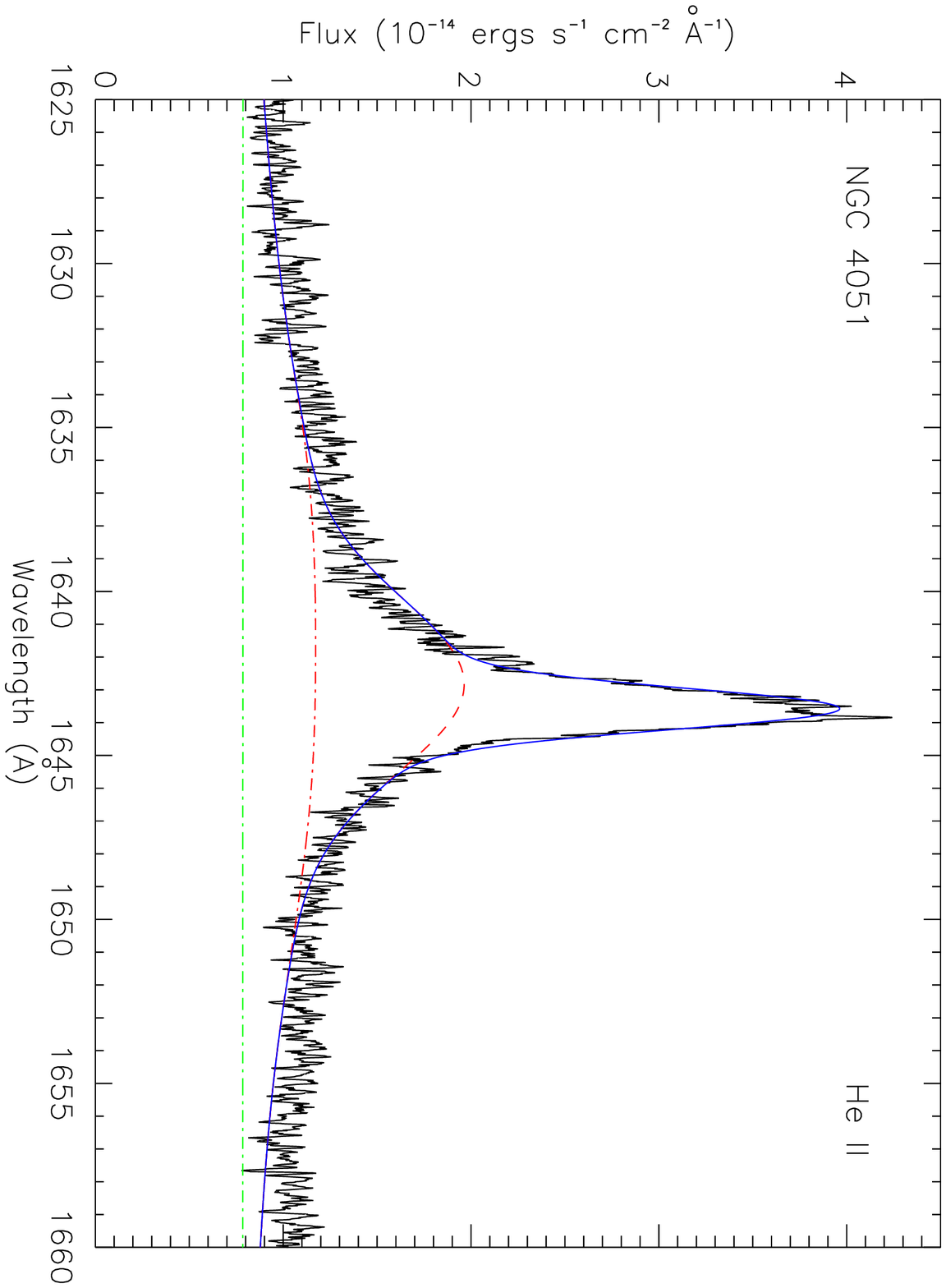]{COS Spectrum of the He~II $\lambda$1640 region and
continuum plus emission-line fits. Components of the fits are: continuum
(dotted-dashed green line), continuum + BLR (dotted-dashed red line), continuum
+ BLR + ILR (dashed red line), and continuum + BLR + ILR + NLR (solid blue
line).}

\figcaption[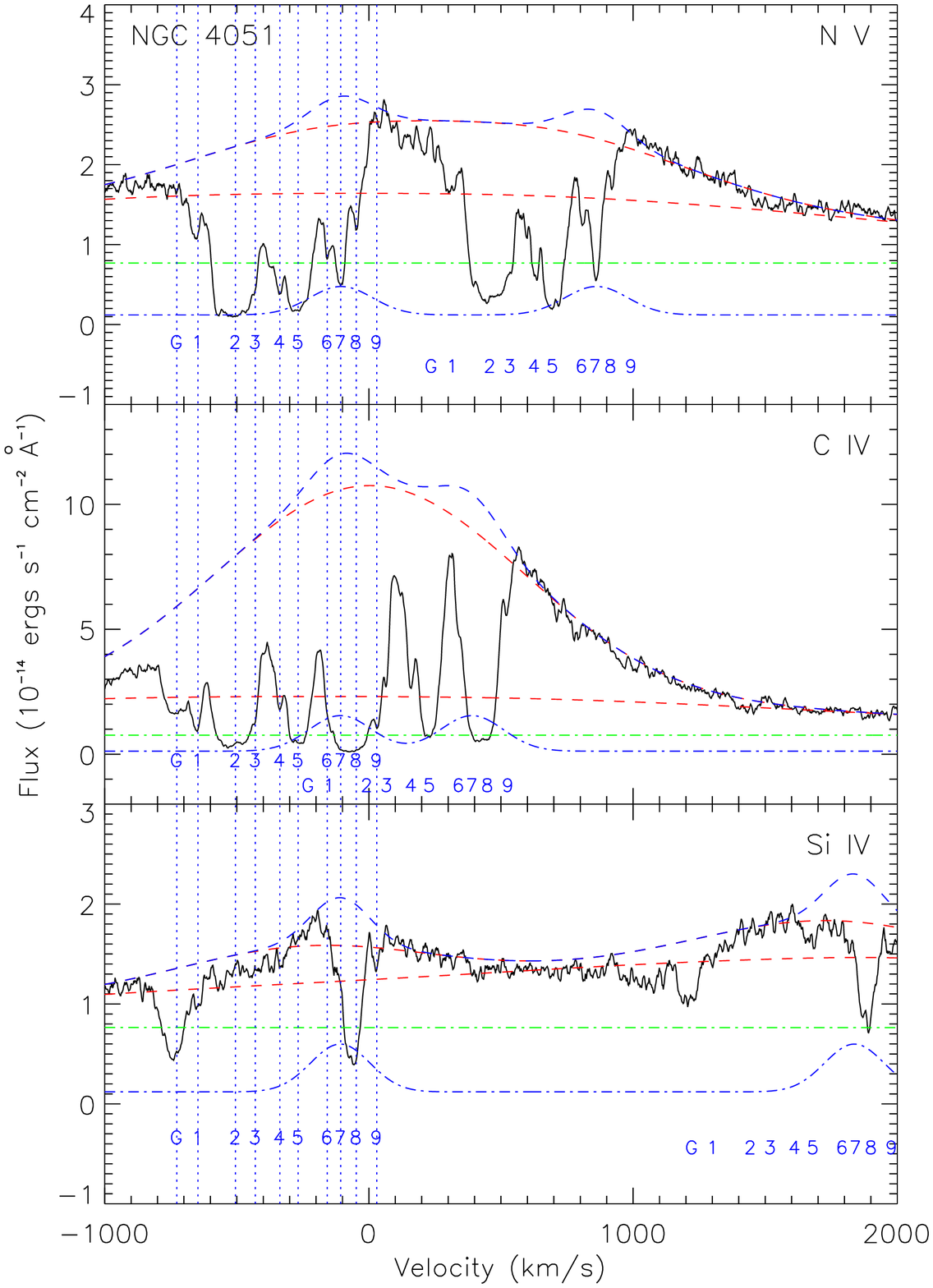]{Intrinsic absorption components in the COS spectrum as in
Figure 3 and continuum plus emission-line fits. Components of the fits are:
continuum (dotted-dashed green line), continuum + BLR (lower dashed red line),
continuum + BLR + ILR (upper dashed red line), and continuum + BLR + ILR + NLR
(dashed blue line). The dotted-dashed blue line shows a model for the uncovered
emission in the intrinsic absorption components, which consists of the excess UV
continuum flux discussed in the text plus the NLR emission.}

\figcaption[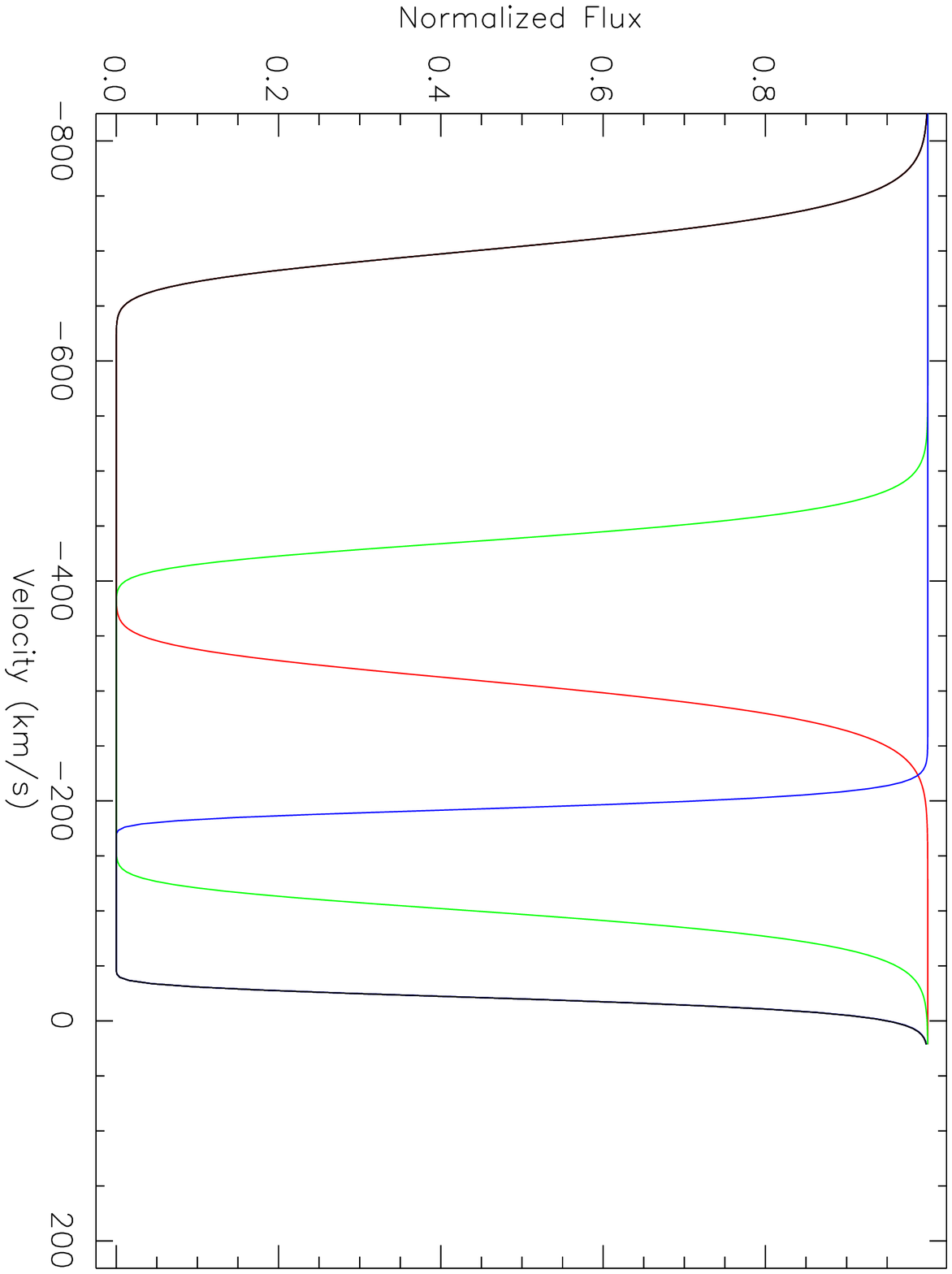]{Separate and combined normalized fluxes for components
2 (red), 5 (green), and 7 (blue), plotted against velocity. The broad, square-troughed combined profile
(shown in black)
resembles the O~VI $\lambda$1032 profile from Kaspi et al. (2004) and
spans the same range in velocity.}

\figcaption[f9.ps]{Comparison of the N~V $\lambda$1240 and C~IV $\lambda$1550
regions, from the COS spectrum, showing the velocities of the individual UV
absorption components, with the O~VIII 1s-2p (18.96 \AA) and O~VII 1s-2p (21.60
\AA) absorption profiles. The oxygen profiles were generated from the {\it
Chandra}/MEG data, with the O~VII binned up by a factor of 2, relative to
O~VIII, due to the lower S/N at those energies. The oxygen profiles have been
shifted by $-$200 km s$^{-1}$ (see discussion in text). The strongest O~VII
absorption occurs near the velocities of UV components 2 and 7/8, while that of
O~VIII is coincident with UV component 2.}

\figcaption[f10.ps]{Simulation of components 6 (blue solid) and 7 (red solid),
assuming $\tau = 10$ in their cores, binned as follows: black solid- total
profile; black dotted - binned to 50 km s$^{-1}$ resolution; black dashed -
smoothed to {\it Chandra}/LETG resolution of 300 km s$^{-1}$. The latter gives
a FWHM $= 250$ km s$^{-1}$, similar to zone 4 in Steenbrugge et al.(2009), while
the unsmoothed is 160 km s$^{-1}$}.

\newpage
\begin{deluxetable}{ccc}
\tablecolumns{3}
\footnotesize
\tablecaption{COS High-Resolution Spectra of NGC 4051}
\tablewidth{0pt}
\tablehead{
\colhead{Grating} & \colhead{Wavelength Coverages} & \colhead{Exposure Time} \\
\colhead{} & \colhead{(\AA)} & \colhead{(s)}
}
\startdata
G130M & 1137 -- 1274, 1291 -- 1429 & 1832 \\
G130M & 1156 -- 1293, 1310 -- 1449 & 2771 \\
G160M & 1389 -- 1558, 1574 -- 1748 & 2771 \\
G160M & 1412 -- 1580, 1601 -- 1772 & 2771 \\
\enddata
\end{deluxetable}

\newpage
\begin{deluxetable}{ccl}
\tablecolumns{3}
\footnotesize
\tablecaption{Kinematic Absorption Components}
\tablewidth{0pt}
\tablehead{
\colhead{Component} & \colhead{$v_r$} & \colhead{FWHM}\\
\colhead{} & \colhead{(km s$^{-1}$)} & \colhead{(km s$^{-1}$)}
}
\startdata
G  & $-$727  & 74\\
1  & $-$647  & 40 \\
2  & $-$505  & 165\\
3  & $-$430  & 63\\
4  & $-$337  & 52 \\
5  & $-$268  & 133 \\
6  & $-$158  & 45 \\
7  & $-$107  & 64 \\
8$^a$  & $-$48  &84 \\ 
9  & $+$30   & 23 \\
\enddata
\tablenotetext{a}{Collinge et al. (2001) call the low-ionization gas at
slightly more negative velocities Component 10.}
\end{deluxetable}

\begin{deluxetable}{lccccc}
\tablecolumns{6}
\footnotesize
\tablecaption{Ionic Column Densities in NGC~4051 (10$^{14}$ cm$^{-2}$)}
\tablewidth{0pt}
\tablehead{
\colhead{} &\multicolumn{5}{c}{Component}\\
\colhead{$N_{ion}$} & \colhead{2} & \colhead{5} &
\colhead{7} & \colhead{8} & \colhead{9}
}
\startdata
H~I    &13$\pm$5      &45$\pm$13    & &$>$960$^a$       &  \\
N~V    &$>$13.7       &$>$10.6      &$>$3.4    &0.28$\pm$0.09  &$<$0.1 \\
C~IV   &$>$7.5        &$>$7.3       &          &$>$9.9$^b$     &0.16$\pm$0.04\\
C~III  &0.3$\pm$0.1:  &0.3$\pm$0.1  & &$>$2.5$^a$       &  \\
C~II   &---    &---   &             &$>$4.8$^b$         &0.20$\pm$0.05\\
Si IV  &$<$0.1        & $<$0.1      &$<$0.1    &0.75$\pm$0.13  &0.05$\pm$0.02\\
Si~III &---    &---   &             &$>$0.46$^b$        &0.019$\pm$0.005 \\
Si~II  &---    &---   &             &$>$0.44$^b$        &0.04$\pm$0.01\\
Fe~II  &---    &---   &             &0.49$\pm$0.21$^a$  &$<$0.21\\
\enddata
\tablenotetext{a}{Includes contributions from Component 7 and 9}
\tablenotetext{b}{Includes contributions from Components 7}
\end{deluxetable}

\begin{deluxetable}{lcccc}
\tablecolumns{5}
\footnotesize
\tablecaption{Photoionization Model Parameters}
\tablewidth{0pt}
\tablehead{
\colhead{Component} & \colhead{logU$^{a}$} & \colhead{log$N_{\rm H}$$^{a}$}
& \colhead{radial distance(pc)} & \colhead{log$n_{\rm H}$$^{a}$}  } 
\startdata
2 & $-0.72$ & 20.17 & 0.50$^{b}$ & 5.57 \\
5 & $-0.64$ & 20.40 & $<$0.09$^{b}$ & $> $6.90\\
7 & $-0.80$ & 20.18 & $>$0.29$^{b}$& $< $6.13\\
8H$^{c}$ & $-2.40$ & 19.0 & $>$ 1.25 $\times 10^{4}$ & $< -1.52$\\
8L$^{c}$ & $-3.50$ & 18.8 & $>$ 1.25 $\times 10^{4}$ & $< -0.44$\\
9$^{c}$ & $-2.78$ & 17.7 &  $>$ 1.93 $\times 10^{4}$ & $< -0.44$ \\
\enddata
\tablenotetext{a}{$N_{\rm H}$ in units of cm$^{-2}$; $n_{\rm H}$ in units of cm$^{-3}$.
Note, while the values of $U$ and $N_{\rm H}$ are those that yield the
column densities listed in Table 5, the uncertainties in these parameters
are an order of magnitude greater than the least significant figures quoted (see discussion
in Section 4.2.
}
\tablenotetext{b}{based on constraints from Steenbrugge et al. (2009).}
\tablenotetext{c}{see discussion in text.}
\end{deluxetable}

\begin{deluxetable}{lcccccc}
\tablecolumns{7}
\footnotesize
\tablecaption{Predicted Ionic Column Densities$^{a}$ (10$^{14}$ cm$^{-2}$)}
\tablewidth{0pt}
\tablehead{
\colhead{} &\multicolumn{6}{c}{Component}\\
\colhead{$N_{ion}$} & \colhead{2} & \colhead{5} &
\colhead{7} & \colhead{8H} & \colhead{8L}& \colhead{9}
}
\startdata
H~I    &26.2     &36.6   &32.6      & 149 &1.44$\times10^{3}$  &19.9 \\
O~VIII  &23.0 &52.5 &17.3  &--- &--- &---\\
O~VII   &411 &764 &375  &--- &--- &---\\
O~VI    &198 &299 &224  &0.19 &--- &---\\
C~IV   &6.62        &8.39      & 9.14        &7.21  &0.10     &0.16\\
C~III  &0.40  &0.43  & 0.64 & 16.1  &9.62    &1.03  \\
C~II   &---    &---   &---               &0.79 &7.21     &0.13\\
N~V    &10.1       &13.8     &12.8    &0.33 &---  &--- \\
Si IV  &---        &---      &---    &1.09 &0.09 &0.05\\
Si~III &---    &---   & ---             &0.79 &0.52     &0.81 \\
Si~II  &---    &---   & ---              &0.06 &1.43    &0.02\\
Fe~II  &---    &---   & ---             &--- &0.27 &---\\
\enddata
\tablenotetext{a}{Blanks indicate predicted $N_{ion}$ $<$ 10$^{12}$ cm$^{-2}$.}
\end{deluxetable}

\begin{deluxetable}{lccccl}
\tablecolumns{6}
\footnotesize
\tablecaption{Comparison of X-ray and UV components}
\tablewidth{0pt}
\tablehead{
\colhead{Component$^{a}$} & \colhead{$v_r$} & \colhead{FWHM} & \colhead{logU} & \colhead{log$N_{\rm H}^{b}$} & \colhead {UV absorber}\\
\colhead{} & \colhead{(km s$^{-1}$)} & \colhead{(km s$^{-1}$)} & \colhead{} & \colhead{} & \colhead {}} 
\startdata
L3b & $-620$ & 467 & 0.56 & 20.7 & comp 1?\\
S3 & $-600$ & 90 & 0.86 & 20.8 & comp 1?\\
L4 & $-510$ & 467 & 1.57 & 21.4 & comp 2?\\
L3a & $-350$ & 467 & 0.76 & 21.0 & comp 4?\\
S2 & $-270$ & 170 & $-0.64$ & 20.4 & comp 5\\ 
S1 & $-210$ & 300 & $-1.2$ & 20.1 & comp 6?, 7\\
L2 & $-20$ & 1175 & $-0.8$ & 20.1 & comp 6?, 7\\
L1 & $+20$ & 467 & $-2.25$ & 20.5 & comp 8, 9?\\
\enddata
\tablenotetext{a}{Model components ``L'' from Lobban et al. (2011) and ``S'' from Steenbrugge et al. (2009)
(see Section 5.1). Values of $v_r$ for Lobban et al. are shifted by $+$200 km s$^{-1}$.}
\tablenotetext{b}{$N_{\rm H}$ in units of cm$^{-2}$.}
\end{deluxetable}

\clearpage
\begin{figure}
\plotone{f1.eps}
\\Fig.~1.
\end{figure}

\clearpage
\begin{figure}
\plotone{f2.eps}
\\Fig.~2.
\end{figure}

\clearpage
\begin{figure}
\plotone{f3.eps}
\\Fig.~3.
\end{figure}

\clearpage
\begin{figure}
\plotone{f4.eps}
\\Fig.~4.
\end{figure}

\clearpage
\begin{figure}
\plotone{f5.eps}
\\Fig.~5.
\end{figure}

\clearpage
\begin{figure}
\plotone{f6.eps}
\\Fig.~6.
\end{figure}

\clearpage
\begin{figure}
\plotone{f7.eps}
\\Fig.~7.
\end{figure}

\clearpage
\begin{figure}
\plotone{f8.ps}
\\Fig.~8.
\end{figure}

\clearpage
\begin{figure}
\plotone{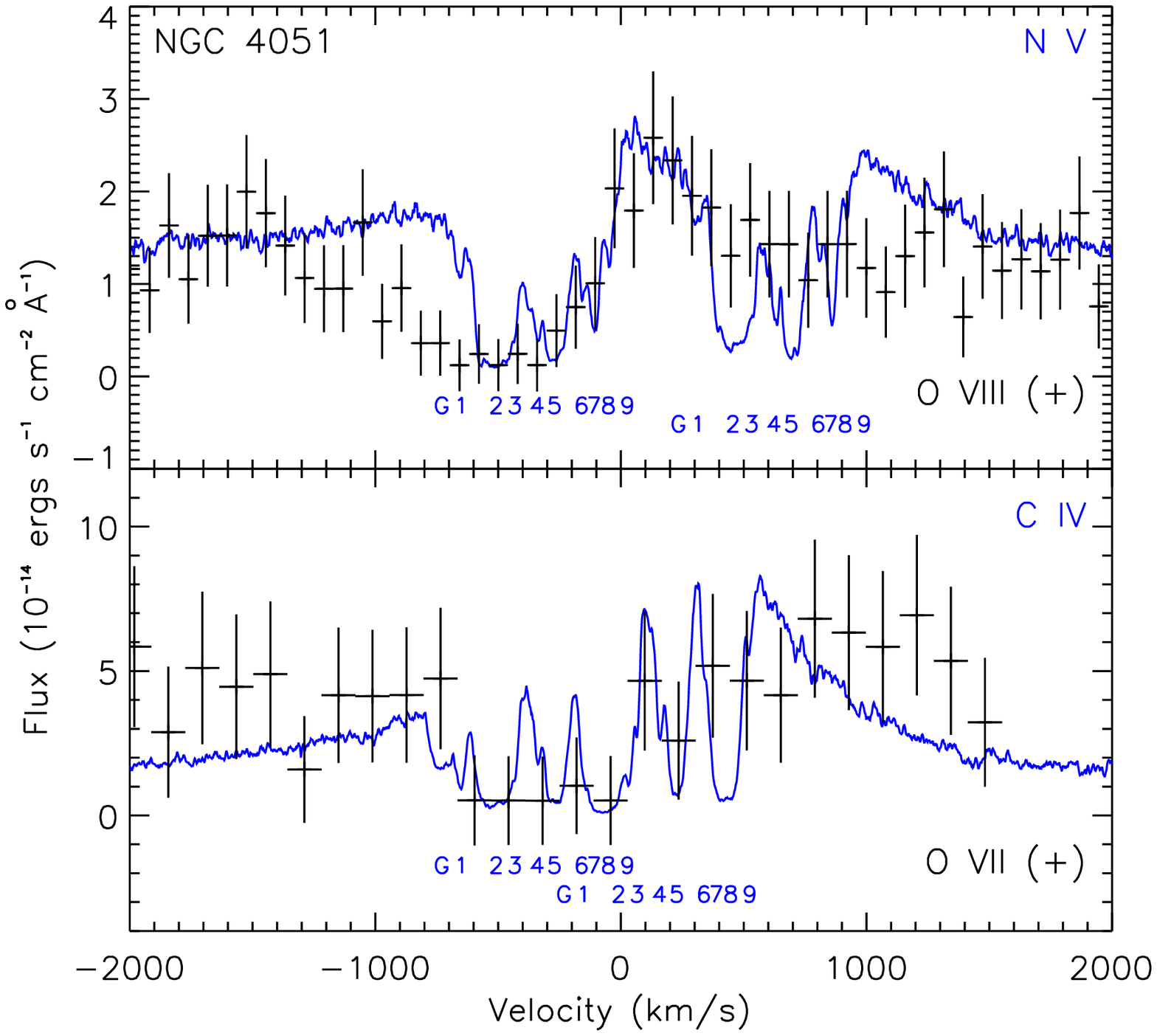}
\\Fig.~9.
\end{figure}

\clearpage
\begin{figure}
\plotone{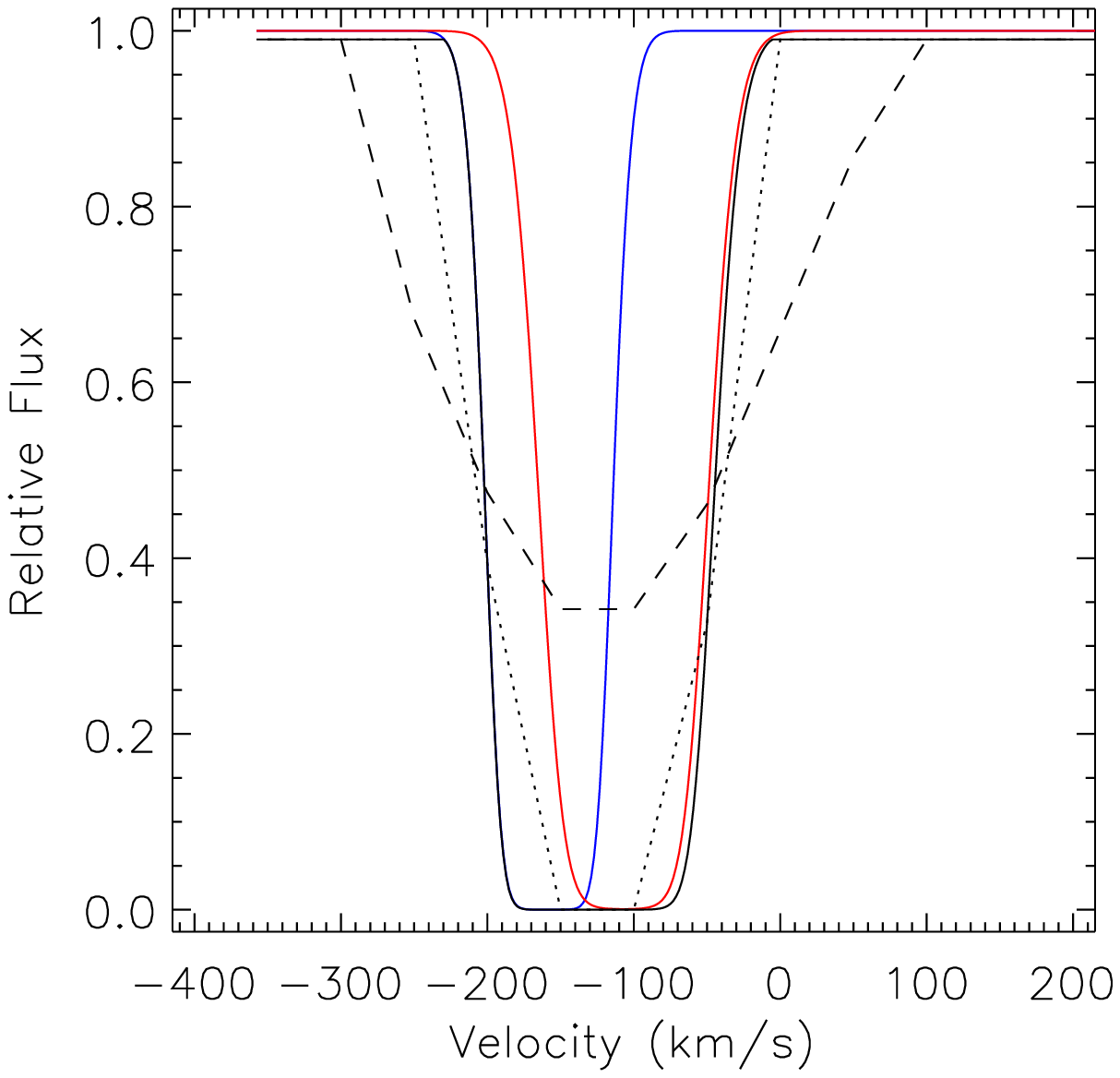}
\\Fig.~10.
\end{figure}
\end{document}